\documentclass[preprint,aps,amsmath,amssymb,showpacs,nofootinbib]{revtex4}
\usepackage{graphicx}
\usepackage{bm}

\newcommand{\cs}{{\bf cs}}

\newcommand{\br}{{\bf r}}
\newcommand{\bp}{{\bf p}}
\newcommand{\bk}{{\bf k}}

\newcommand{\df}{\stackrel{\rm def}{=}}

\newcommand{\Vsc}{V_{\rm sc}}

\newcommand{\Det}{\operatorname{Det}}

\newcommand{\smeq}{\! = \!}

\newcommand{\kf}{k_{\scriptscriptstyle F}}
\newcommand{\vf}{v_{\scriptscriptstyle F}}
\newcommand{\pf}{p_{\scriptscriptstyle F}}
\newcommand{\Ef}{E_{\scriptscriptstyle F}}
\newcommand{\lambdaf}{\lambda_{\scriptscriptstyle F}}

\newcommand{\PSImagx}[2]{\includegraphics[width=#2]{#1}}

\newcommand{\Nsm}{{ \Delta N}} 
 
\newcommand{\Si}{{ {\cal S}_i }} 

\newcommand{\varSi}{{{\rm Var}[{\cal S}_i]  }}

\begin{document}

\title{Semiclassical theory of non-local statistical measures: 
residual Coulomb interactions} 
\author{Denis Ullmo$^1$, Steven Tomsovic$^{1,2,3}$,  and Arnd B\"acker$^4$}
\affiliation{$^1$CNRS; Univ.~Paris-Sud; LPTMS UMR 8626; 91405 Orsay Cedex, France} 
\affiliation{$^2$Max-Planck-Institut f\"ur Physik komplexer Systeme; D-01187 Dresden, Germany} 
\affiliation{$^3$Department of Physics and Astronomy; Washington State University; Pullman, WA 99164-2814 (permanent address)} 
\affiliation{$^4$Institut f\"ur Theoretische Physik; Technische Universit\"at Dresden; 01062 Dresden, Germany} 
\date{\today}

\begin{abstract}

In a recent letter [Phys.~Rev.~Lett.~{\bf 100}, 164101 (2008)] and within the context of quantized chaotic billiards, random plane wave and semiclassical theoretical approaches were applied to an example of a relatively new class of statistical measures, i.e.~measures involving both complete spatial integration and energy summation as essential ingredients.  A quintessential example comes from the desire to understand the short-range approximation to the first order ground state contribution of the residual Coulomb interaction.   Billiards, fully chaotic or otherwise, provide an ideal class of systems on which to focus as they have proven to be successful in modeling the single particle properties of a Landau-Fermi liquid in typical mesoscopic systems, i.e.~closed or nearly closed quantum dots.  It happens that both theoretical approaches give fully consistent results for measure averages, but that somewhat surprisingly for fully chaotic systems the semiclassical theory gives a much improved approximation for the fluctuations.  Comparison of the theories highlights a couple of key shortcomings inherent in the random plane wave approach.  This paper contains a complete account of the theoretical approaches, elucidates the two shortcomings of the oft-relied-upon random plane wave approach, and treats non-fully chaotic systems as well.

\pacs{03.65.Sq, 05.45.Mt, 71.10.Ay, 73.21.La, 03.75.Ss}
\end{abstract}

\maketitle


\vfill\eject

\newcommand{\AB}[1]{{\scriptsize\tt{AB: #1}}}

\section{Introduction}
\label{sec:introduction}

Finite and low-dimensional quantum systems often possess statistical properties whose deviations from  universality contain some basic dynamical information about the system~\cite{Bohigas93,Aleiner97,Mirlin00,Schomerus05}.  A recurring challenge is to understand precisely what information is buried in those statistical deviations and how to extract it.  Before that can be addressed, the universal behaviors themselves must be understood.  At the heart of many such universalities generally lurks a connection to the Bohigas-Giannoni-Schmit (BGS) conjecture~\cite{Bohigas84,Bohigas91}, which asserts that systems with underlying chaotic dynamics have the fluctuation properties found in random matrix theory~\cite{Mehta04}.  If interest lies in quantities for which the position representation of the eigenfunctions is critical, a random plane wave model is often introduced that augments the BGS conjecture~\cite{Berry77, Voros79}.  The primary goal of extracting system specific information can then proceed, but generally requires a more powerful theory.

The preponderance of statistical measures heretofore introduced for analysis are local in energy,  configuration space, or both.  Examples are given by the Dyson-Mehta cluster functions~\cite{Mehta04}, and the amplitude distribution and short-range two-point correlation function $c(|\br - \br'|) = \langle \psi(\br) \psi(\br') \rangle$ of a given eigenfunction, $\psi(\br)$.   On the other hand, there has been a rather recent introduction of new, non-local statistical measures~\cite{Ullmo03prl, Miller05, Garcia08}.  They have been motivated by the need to understand the interplay between interferences and interactions in mesoscopic systems.  For example, one place where this interplay is known to have an important role is the addition spectra of quantum dots; other examples are coming from cold fermionic gasses.  Focussing on just one motivation, the addition spectrum is experimentally accessible through
the position of the conductance peaks in Coulomb Blockade transport
measurements \cite{Sivan96, Simmel97, Patel98a, Simmel99, Luscher01}.
Fluctuations of peak spacings are associated with interference
effects, and experimentally are clearly incompatible with a
non-interacting description of the conduction electrons in the dots.
Indeed, a non-interacting description predicts a strong bimodality of
the peak spacing distribution -- associated with an odd-even character of the
number of (spin-$1/2$) electrons -- and this is not observed.
Assuming further that the dot possesses a chaotic dynamics, the
distribution for odd-$N$ spacings should show a characteristic Wigner
surmise shape, whereas a density similar to a Gaussian with
extended tails is observed.

At first, it was argued that a non-Fermi liquid description of the
interacting electrons might be necessary to interpret the experimental
data.  However, the picture which has now emerged \cite{Blanter97,
  Ullmo01prbb, Usaj01, Usaj02, Aleiner02PhysRep} is that although an
understanding of both peak spacings and ground state spin
distributions is still incomplete, it is reasonable to expect
that most phenomena will eventually be explained within a Fermi
liquid framework.  More specifically, the electrons can be thought of
as quasi-particles confined by a potential $U_{\rm conf}(\br)$, which
could in practice be computed within a self-consistent
Thomas-Fermi-like approximation \cite{Ullmo04prb}.  They also interact
weakly through a screened Coulomb interaction $\Vsc(\br,\br')$.  For
dots significantly larger than the screening length, confinement does
not modify appreciably the screening process, and the bulk expression
for $\Vsc(\br-\br')$ is appropriate.  Furthermore, for the
experimentally relevant gas parameter $r_s$ being of order one, the
screening length is not much different from the Fermi wavelength
$\lambdaf$.  Under these circumstances, the problem is well described
by the short range approximation
\begin{equation} \label{eq:Vsr}
\Vsc(\br-\br') = \frac{F_0^a}{\nu} \delta(\br-\br') \; ,
\end{equation}
with $\nu$ the mean local density of states (including the spin
degeneracy) ($\nu \smeq m/\pi\hbar^2$ for $d \smeq 2$)
and $F_0^a$ the dimensionless Fermi liquid parameter
\cite{PinesNozieresVol1} (for $d \smeq 2$ and $r_s$ of order one,
$F_0^a$ is in the range 0.6-0.8).  Boundaries could potentially modify
this picture somewhat, but it is expected that a slightly modified
$F_0^a$ would be sufficient to capture the effects; since this is not
the focus of our study, we leave it for future consideration.   

In this approximation, the first order contribution of the residual
interactions to the ground 
state energy can be expressed in the simple form
\begin{equation} 
\label{eq:ERI}
\delta E^{\rm RI} = \frac{ F_0^a }{\nu} \int d\br \, n_\uparrow(\br)
n_\downarrow(\br) \; ,
\end{equation}
with $n_\sigma$ the unperturbed ground state density of particles with
spin $\sigma$.  This expression was the starting point for a study
demonstrating the increased importance of $\delta E^{\rm RI}$ if the
dynamics are not fully chaotic~\cite{Ullmo03prl}. 

Because $n_\sigma(\br)$ can be expressed as a sum over the absolute
square of single particle eigenfunctions which are occupied, the
mesoscopic fluctuations of the residual energy term $\delta E^{\rm
  RI}$ of Eq.~(\ref{eq:ERI}), or of similar quantities, are related to
the fluctuations of the one-particle eigenstates of the
unperturbed system.  However, in contrast to the correlation function
$c(|\br - \br'|)$, Eq.~(\ref{eq:ERI}) involves both an integration over space and
summation over energy, and therefore in this way, it is probing new
aspects of the fluctuation properties of the eigenstates.

The goal of this paper is to follow up on our recent work~\cite{Tomsovic08} on the average and fluctuating parts of the quantity in Eq.~(\ref{eq:ERI}) in several ways.   To begin with a complete account is given of two theoretical approaches using random plane waves and semiclassical theory.  As already shown, for fully chaotic quantized billiards the two methods give identical leading functional dependence on wave vector and system size for average and fluctuation properties.  In addition, the significant differences between Dirichlet and Neumann boundary conditions, can be understood.   On the other hand, in the statistical limit of uniformity, the fluctuation properties differ in two ways in the prefactor.   The semiclassical treatment in the spirit of the Gutzwiller trace formula~\cite{Gutzwiller71,Gutzwiller90} helps identify dynamical correlations and a term missing from the expressions derived with the random plane wave model.  Thus, the Gutzwiller periodic orbit approach provides both a deeper understanding of the mechanism underlying the fluctuations and a good quantitative agreement with exact numerical calculations for the examples considered.  Next, non-fully chaotic systems are partially addressed.  The fluctuations are found to be greatly magnified in two essential ways.  One relates to non-uniform projected classical densities and the other to enhanced dependence on wave vector and system size.  

The organization of the paper is as follows.  In Sect.~II, the necessary background material and notations are introduced, including a more precise definition of the statistical quantities to be studied.
Also introduced in this section is the random plane wave model, which is then used to analyze in Sect.~III the mean and fluctuating behaviors.  The random plane wave modeling reproduces quite accurately the mean behaviors, but it fails by most of an order of magnitude to predict quantitatively the fluctuations.  This motivates the introduction in Sect.~IV of the semiclassical analysis in
terms of classical trajectories.  This approach corrects the random plane wave method overestimate of the fluctuations.  In Sect.~V, extended semiclassical methods are introduced for systems which are not fully chaotic.  Finally Sect.~VI contains a discussion and summary.

\section{Preliminary considerations}
\label{sec:background}

To  begin, it is worth motivating the introduction of a non-local statistical measure in a little more detail.  For example, consider two-degree-of-freedom chaotic quantized billiard systems with
one body eigenstates $\psi_i(\br)$ and energies $E_i$.  In the
absence of interactions, the many-body eigenstates are Slater
determinants characterized by spin-dependent occupation numbers
$f_{i,\sigma}=0$ or $1$.  Within the short range approximation,
Eq.~(\ref{eq:Vsr}), the contribution of the residual interactions to
the ground state can be written in first order perturbation theory as
\begin{eqnarray} 
\label{eq:ERI2}
\delta E^{\rm RI} &=&  \frac{F_0^a\Delta}{2}\sum_{i,j} 
f_{i,(+)} f_{j,(-)}  M_{ij}
\end{eqnarray}
where $\Delta$ is the mean single particle level spacing in the neighborhood of the Fermi surface.
$M_{ij}$ is given by
\begin{equation} 
\label{eq:Mij}
M_{ij} =  {\cal A} \int d\br \left| \psi_i(\br) \right|^2 \left| \psi_{j}(\br)
\right|^2  \; ,
\end{equation}
where $\cal A$ is the area of the billiard.  With this definition, the
$\{M_{ij}\}$ are dimensionless quantities with a mean value expected
to be roughly equal to unity for $i\ne j$, and to three for $i=j$.
These expectations would apply to uncorrelated Gaussian random
amplitudes assuming time reversal invariance holds; ahead more precise
results are derived.

Of main interest is how the value of $\delta E^{\rm RI} $ changes when
a particle is promoted from one orbital to another (as opposed to the
variations of the residual energy as particles are added into the
system).  Consider that the ground
state of the non-interacting $N$-particle system is such that
the levels below the Fermi energy are doubly occupied except for the
last level $i_F$, which may be singly or doubly occupied depending on
the parity of $N$.  Promoting a particle from the orbital $i_F$ to
$i_F+1$ has a one-particle energy cost $(E_{i_F +1} -
E_{i_F})$.  If however this is compensated by the
corresponding difference in residual energy, the ground state
occupation numbers $f_{i,(\pm)}$ will be modified by the interactions,
yielding in some circumstances non trivial, i.e. different
from $0$ or $1/2$, 
ground state spins.  Imagining $\{f_{i,(+)} M_{ij} f_{j,(-)}\}$
in the form of a (square or nearly square) symmetric matrix, shifting
an occupancy from one level to another subtracts the column or row
being vacated and adds a column or row to the newly occupied orbital
(row or column depends on the spins of the removed and added
particles).  Apart from a couple individual $M_{ij}$'s near the diagonal, the
difference in the residual interaction can therefore be expressed in terms
of the difference of two sums of the form
\begin{equation}
\label{eq:si}
{\cal S}_i = \sum_{j=1}^i M_{ij}  \; .
\end{equation}
As the $\{M_{ij}\}$ are positive definite, $ {\cal S}_i$ has a locally
defined, increasing-with-index, positive mean.  However, because one
column or row is added and another subtracted, their means largely
cancel, and thus the mean $ {\cal S}_i$ behavior cannot be involved in
altering the ground state occupancies of the single particle levels
defined by $U_{\rm conf}(\br)$. [The mean of the few individual
$M_{ij}$'s near the diagonal not included in ${\cal S}_i$ might
though.]  However, if the fluctuations of 
${\cal S}_i$ or $\{M_{ij}\}$ 
are sufficiently large, they have the potential to alter the
nature of the ground state.  Thus, statistical measures based on the properties of the ${\cal S}_i$ are of fundamental interest, in particular their mean values and fluctuations.

The ${\cal S}_i$ have the unusual character that they are integrated over space and involve a sum over eigenenergies.  Their investigation thus requires two basic ingredients.  With the
definition
\begin{equation}
\label{eq:nr}
N(\br ; E) \equiv \int_0^E {\rm d}E^\prime n(\br; E^\prime) =
\sum_{i=1}^\infty \left|\Psi_i(\br)\right|^2 
\theta \left( E - E_i \right) 
\end{equation}
${\cal S}_i$ can be expressed as
\begin{equation}
\label{eq:sinr}
{\cal S}_i = {\cal A} \int {\rm d}\br\ \left|\Psi_i(\br)\right|^2
\sum_{j\le i}  \left|\Psi_j(\br)\right|^2 = {\cal A}\int {\rm d}\br\
\left|\Psi_i(\br) \right|^2N(\br;E_i^+) \; ,
\end{equation}
with the understanding that $E_i < E_i^+ < E_{i+1}$ (and assuming for simplicity that there are no degeneracies).  One of the two
required ingredients is the behavior of $N(\br;E)$.  As it results
from a summation over the absolute square of eigenstates up to a
certain energy, it is dominated by a secular behavior; see Fig.~13 in
Ref.~\cite{Baecker98} for example.  The secular component $N_{\rm  sec}(\br;E)$ emerges from an 
energy smoothing which, although local, is also necessarily broader than the Thouless energy~\cite{Ullmo08}; ahead the notation $\langle \cdot \rangle$ is introduced to denote this averaging.  This energy smoothing implies that only dynamics on a time scale shorter than the shortest periodic orbit is relevant, and thus this decomposition is independent of whether the system dynamics is regular, fully chaotic or has some other character.  $N_{\rm  sec}(\br;E)$  has been shown to be given by an excellent semiclassical (asymptotic) approximation \cite{Hoernander85,Baecker98}
\begin{equation}
\label{eq:nre}
 N_{\rm sec}(\br;E) = \frac{N_W(E)}{{\cal A}}\left[ 1 \pm \frac{J_1(2kx)}{kx}
 \right]  
\end{equation}
where the coordinate $x$ is defined locally as the perpendicular
distance from the boundary, $k$ is the magnitude of the wave vector at
energy $E$, and the $+$ sign is for Neumann
boundary conditions and the $-$ sign for Dirichlet.  
Here $N_W(E)$ refers to just the leading term of the Weyl formula, 
$N_W(E) = \frac{m{\cal A}}{2\pi \hbar^2} E$.
The validity is governed by $kL\gg1$, where $L$ is a length scale, 
specified ahead in the paper, but which necessarily must be shorter
 than the width of the system.  Away from the
boundary, the secular behavior approaches an overall constant.
However, the existence of boundary conditions and a minimum wavelength
scale combine to create persistent oscillations (Friedel oscillations), 
which are maximal near the boundary and which fade away
toward the interior of the billiard.  The negative sign of the Bessel
function for Dirichlet boundary conditions respects the vanishing of
eigenfunctions at the boundary as it must.  Note that just as the critical portion of the density of states can be expressed as a series with volume, boundary, curvature, and oscillatory components, the same is true of $ N(\br;E)$.  The above expression does not include the curvature components and must therefore be missing at least part of the $O([kL]^{-2})$ corrections.

As noted $N_{\rm sec}(\br;E)$ is a smooth function of the parameter
$E$, but ${\cal S}_i$ actually involves $N_{\rm sec}(\br;E_i^+)$,
which changes abruptly at the points where the energy surpasses each
eigenvalue (i.e.~is a function of $i$).  The former quantity does not
have a monotonous dependence in the number of particles $i$ since
$E_i$ contains the Gutzwiller corrections from periodic orbit theory
\cite{Gutzwiller71,Gutzwiller90} that determine the precise positions
of the levels.  We therefore consider instead the slightly modified
and properly normalized decomposition
\begin{eqnarray}
\label{eq:nre2}
 N(\br;E_i^+) &=& N_{\rm sec}(\br;E_i^+) +\delta N(\br;E_i^+) \nonumber \\
 N_{\rm sec}(\br;E_i^+) &\approx & \frac{i}{{\cal A} \left( 1 \pm \frac{{\cal L} }{k_i {\cal
         A}} \right) }\left[ 1 \pm \frac{J_1(2k_i r) }{ k_i r} \right] \; ,
\end{eqnarray}
where the $\cal L$ is the billiard perimeter and not to be confused
with length scale $L$ mentioned above.  This decomposition has the
further advantage that the fluctuations $\delta N(\br;E^+_i)$ not
contained in the secular behavior average to zero when integrated over
space.  In this way, density of states oscillations, which are not of
interest here, do not get intertwined with the fluctuations that are
the focus of this study.  To the order of corrections incorporated in
Eq.~(\ref{eq:nre2}), $k_i$ can equally be defined as
$\sqrt{2mE_i}/\hbar$ or as the mean value obtained from the Weyl
series.

The other main necessary ingredient is the behavior of $\left|\Psi_i(\br) \right|^2$ which leads to the two principal approaches contained in this paper.   One approach is to rely upon a statistical model which uses an ensemble of random plane waves to mimic the properties of chaotic eigenstates\cite{Berry77,Voros79,Longuet52}  and the other is to use a semiclassical theory building on the work of Bogomolny~\cite{Bogomolny88}.  We begin with the random plane wave modeling as it is technically simpler.

\section{Random plane wave modeling}
\label{sec:rpw}

The random plane wave model \cite{Berry77,Voros79,Longuet52} 
has been introduced in which the eigenstates are represented, in the absence of any symmetry by a
random superposition of plane waves $\sum_l a_l \exp(i
\bk_l \cdot\br)$ with wave-vectors of fixed modulus $|\bk_l|=\kf$
distributed isotropically.  Time reversal invariance may be introduced as a
correlation between time reversed plane waves such that the
eigenfunctions are real.  Similarly, the presence of a planar
boundary imposes a constraint between the coefficients of plane waves
related by a sign change of the normal component of the
wave-vector $\bk_l$.  Near a boundary, and using a system of
coordinates $\br =  \hat {\bf x} +  \hat {\bf y}$ with $\hat {\bf x}$
and $\hat {\bf y}$ the vectors respectively  perpendicular and parallel to the boundary (of norm $x$ and norm $y$), eigenfunctions are therefore locally
mimicked statistically by a superposition,
\begin{equation}
\label{eq:rpwm}
\psi_i(\br) = \frac{1}{N_{\rm eff} }\sum_{l=1}^{N_{\rm eff}}
a_l \cs \left( 
  {\bf k}_l \cdot \hat {\bf x} \right) 
   \cos\left( {\bf k}_l\cdot \hat{\bf y} +
  \varphi_l\right) 
\end{equation}
where $\cs(\cdot) \df \sin(\cdot)$ for Dirichlet and $\cos(\cdot)$ for
Neumann boundary conditions \cite{Ber2002}.  The phase angle
$\varphi_l$, the real amplitude $a_l$, and the orientation
of the wave vector ${\bf k}_l$, are all chosen randomly.  The
amplitudes $a_l$ are zero-centered independent Gaussian random
variables with $\langle a_l a_{l'} \rangle = \delta_{l
  {l'}} \sigma^2$.

To complete the model, it is necessary to determine the variance $\sigma^2$ which is fixed by the
normalization of the wavefunctions.  Here this constraint
is imposed only on average, rather than for each individual state.
A priori, proceeding in this way might be expected to miss weak correlations between the
eigenfunctions.  The question is whether one should expect them  to be insignificant.  In principle, the answer is yes, but only if local properties of the eigenfunctions are being considered
and the effective dimensionality is large as it would be for
$k_FL\gg1$, where $k_F$ is the Fermi wave vector.  This issue is further discussed in the semiclassical theory of Sect.~\ref{sec:gutzwiller} ahead.

Using the random plane wave representation Eq.~(\ref{eq:rpwm}) we have
\begin{equation}
\left|\psi_i(\br)\right|^2 = \frac{1}{N^2_{\rm eff}
}\sum_{l,{l'}=1}^{N_{\rm eff}} 
a_l a_{l'} \cs\left( {\bf k}_l \cdot \hat {\bf x}\right)
\cs\left( {\bf k}_{l'}\cdot \hat {\bf x}\right)
\cos\left( {\bf k}_l \cdot \hat {\bf y} +
  \varphi_l\right)  \cos\left( {\bf k}_{l'} \cdot \hat {\bf y} +
  \varphi_{l'}\right)  \;,
\end{equation} 
whose expectation value is given by 
\begin{eqnarray}
\label{eq:normalize}
 \left<\left|\psi_i(\br)\right|^2\right> &=& \frac{\sigma^2 }{ 4
   N^2_{\rm eff} }\sum_{l=1}^{N_{\rm eff}}\left[ 1\pm \cos\left( 2{\bf
       k}_l \cdot \hat {\bf x}\right) \right] =  \frac{\sigma^2 }{ 4
   N_{\rm eff} } \left[ 1\pm \frac{1}{ 2\pi}\int_0^{2\pi}{\rm d}\theta \,
   \cos\left( 2 \kf  x\cos\theta\right) \right]  \nonumber  \\ 
 &=& \frac{\sigma^2}{ 4 N_{\rm eff} }  \left[ 1\pm  J_0(2\kf x) \right]  \;,
\end{eqnarray}
where we follow the convention that the upper sign refers to Neumann
and the lower sign to Dirichlet boundary conditions respectively.  The ensemble transformation  $\frac{1}{N_{\rm eff}} \sum_l \rightarrow \frac{1}{2\pi} \int_0^{2\pi} d\theta_l $ has been employed to simplify the calculation.   Above, the norms of the wave vectors, equal to $k_i$, are assumed at or near enough the Fermi surface, that they can be denoted by $\kf$.
Integrating over the area of the billiard to fix the normalization
gives 
\begin{eqnarray}
1 &=& \int d\br \, \left<\left|\psi_i(\br)\right|^2\right> =
\frac{ \sigma^2}{ 4N_{\rm eff} }\left[ {\cal A} \pm {\cal L}
  \int_0^\infty {\rm d}r 
  J_0(2\kf x)\right] \nonumber \\ 
&=& \frac{ { \cal A} \sigma^2}{ 4N_{\rm eff} }\left( 1 \pm \frac{{\cal L} }{
    2\kf {\cal A}} \right) 
\end{eqnarray}
which fixes the variance $\sigma^2$ to next to leading order in $\kf L$.

\subsection{Average Properties}
\label{sec:averageproperties}

The first step in calculating the average behavior of ${\cal S}_i$ is to isolate its secular and fluctuating behavior.  The model above  implies
\begin{equation}
\label{eq:envelope}
{\cal A}\left<\left|\psi_i(\br)\right|^2\right> =  \frac{1 }{  \left(
    1\pm \frac{{\cal L} }{ 2\kf {\cal A}} \right)} \left[ 1\pm  J_0(2\kf x)
\right]  \; .
\end{equation}
and that is consistent with $N_{\rm  sec}(\br;E^+_i)$, i.e.~the
Friedel oscillation contributions to $N(\br;E_i^+)$.  This form
applies more generally than the random plane wave model.  Just as
$N_{\rm  sec}(\br;E^+_i)$ is independent of system dynamics, so also
is this result for the same reasons.  For example, it would emerge for
integrable systems as well assuming the averaging is over energy
intervals greater than the Thouless energy.  Finally, note that from
the correction to unity of the leading constant in this expression,
one sees that if the boundary conditions are Dirichlet, the local mean
behavior of $\left<\left|\psi_i(\br)\right|^2\right> $ well into the
interior is slightly elevated above $1/{\cal A}$ to compensate for the
reduced density near the boundary, and just the opposite for Neumann
boundary conditions.      


Let
\begin{eqnarray}
\epsilon_i(\br) &=& {\cal A}\left|\psi_i(\br)\right|^2 -1 =  {\cal
  A}\left<\left|\psi_i(\br)\right|^2\right> - 1 + {\cal
  A}\left[\left|\psi_i(\br)\right|^2  -
  \left<\left|\psi_i(\br)\right|^2\right>\right]   \nonumber \\ 
&=& \pm \frac{ 1 }{  \left( 1\pm \frac{{\cal L}} {2 \kf {\cal A}} \right)}
\left[ J_0(2 \kf x) - \frac{{\cal L} }{ 2 \kf{\cal A}}  \right] + {\cal
  A}\bigl[\left|\psi_i(\br)\right|^2  -
  \left<\left|\psi_i(\br)\right|^2\right>\bigr] \; .\label{eq:psidef}
\end{eqnarray}
Under spatial integration, $\epsilon_i(\br)$ as well as both halves of
the second expression above each separately vanish whereas only the second half
of the expression is affected by taking the expectation value and in
that case it vanishes.  Ahead, this decomposition simplifies the discussion of
the fluctuations.

Returning to the calculation of ${\cal S}_i$, substituting the
relations from Eqs.~(\ref{eq:nre2},\ref{eq:psidef}), integrating
the constant terms,  and dropping second order terms gives 
\begin{equation} \label{eq:Si} {\cal S}_i = i \pm \frac{i }{ {\cal A}
  } \int {\rm d}\br \ \frac{J_1(2 \kf x)}{\kf x} \epsilon_i(\br) \;.
\end{equation}
and therefore
\begin{eqnarray}
\label{eq:simean}
\left<{\cal S}_i \right> &=& i  \pm \frac{i}{ {\cal A}} \int {\rm
  d}\br \  \frac{J_1(2\kf x) }{ \kf x}  \left<\epsilon_i(\br)\right>
\nonumber \\ 
&=& i + \frac{ i}{ {\cal A}} \int {\rm d}\br \  \frac{J_1(2\kf x) }{ \kf x}  J_0(2\kf x) \nonumber \\ 
&=& i \left( 1+  \frac{2{\cal L} }{\pi \kf  {\cal A}}\right) 
\end{eqnarray}
where we choose $\kf =\sqrt{2mE_i}/\hbar$.  Note the first correction
is independent of whether the boundary conditions are Neumann or
Dirichlet (see Appendix~\ref{app:secondorder} for the calculation of
second order terms not related to curvature and discontinuities in the
boundary).  We stress furthermore that Eq.~(\ref{eq:simean}) is 
applicable independently of the nature of the dynamics, and in
particular apply equally well to integrable and chaotic systems. A
simple semiclassical proof of this will be given in
section~\ref{sec:gutzwiller}.

\begin{figure}[t]
\PSImagx{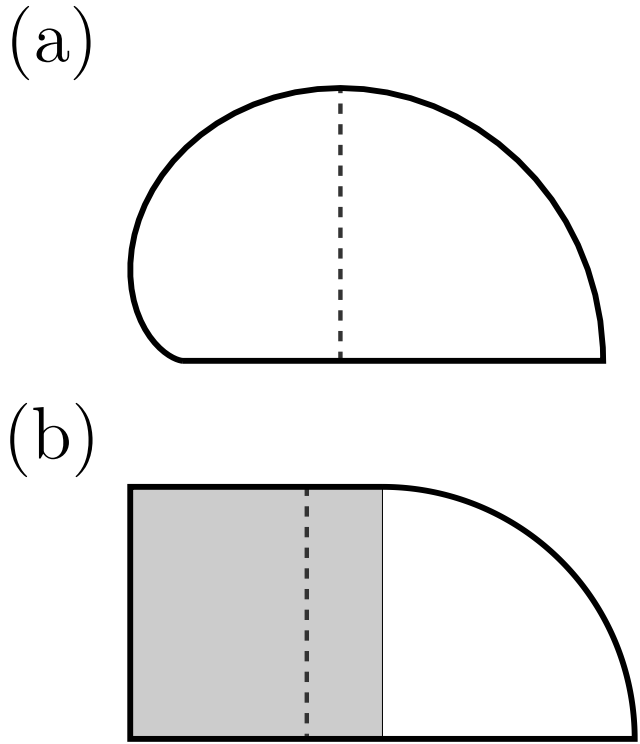}{8.6 cm}
\caption{Drawing of the desymmetrized cardioid and stadium billiard boundaries.  The dashed lines illustrate the paths of the shortest periodic orbits for each system.  Whereas this orbit is isolated in the cardioid billiard, in the stadium billiard it is a member of a continuous one-parameter family of identical orbits,
indicated by the grey-shaded rectangular region.} 
\label{fig:billiards}
\end{figure}

\begin{figure}[t]
\PSImagx{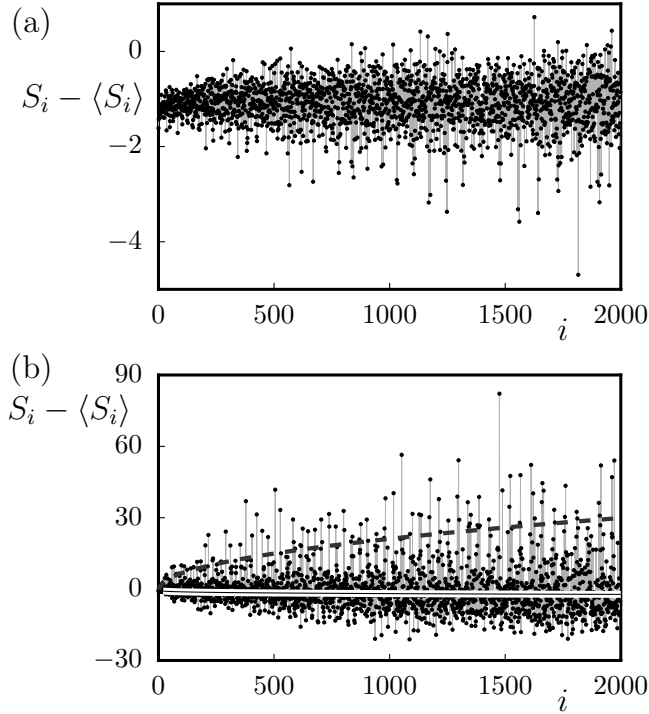}{8.6cm}
\caption{Comparison of ${\cal S}_i$ and the approximation given in the last line of Eq.~(\ref{eq:simean}).  The difference ${\cal S}_i  - \left<{\cal S}_i \right>$ is plotted versus the eigenstate number $i$.  In (a), the results for the odd parity eigenstates of the cardioid billiard using Dirichlet boundary conditions are shown.  In (b), the same results for the even-even eigenstates of the stadium billiard with Dirichlet boundary conditions are shown, except that the mean overall constant term has been numerically subtracted.  Note the significant increase in the scale of the fluctuations about the mean for the stadium billiard.  Ahead in Sect.~\ref{sect:bb}, a rudimentary theory is given for bouncing ball modes that leads to Eqs.~(\ref{eq:resultstad},\ref{eq:negbias}) shown as the dashed and solid lines respectively in (b).} 
\label{fig:mean}
\end{figure}

The well known chaotic cardioid~\cite{Robnik83, Robnik84, Baecker95} and stadium billiards~\cite{Bunimovich74,Bunimovich79,Tomsovic93} are highly suited to illustrating the precision of this relation.    The two symmetry-reduced billiard boundaries are illustrated in Fig.~\ref{fig:billiards} along with their shortest periodic orbits to which we return ahead in the discussion of the Fourier transform of the fluctuations.
Comparison to Eq.~(\ref{eq:simean})  is shown with a computation of ${\cal S}_i$ using the first 2000 odd parity eigenstates of the cardioid billiard and the same number of even-even eigenstates of the stadium billiard; the latter calculation tests the effects of Neumann boundary conditions.  Figure~(\ref{fig:mean}) plots the differences, ${\cal S}_i  - \left<{\cal S}_i \right> $, versus the state index for both 
billiards.   As often happens with semiclassical approximations,  even though the result is asymptotic, it is valid right down to either ground state.  Note that  the second order corrections have not been included in the secular behavior equations, Eqs.~(\ref{eq:nre2}, \ref{eq:envelope}), and so it is seen that the cardioid results are not centered on zero, but on a constant somewhere nearby.  The same is also true for the stadium, except that the mean constant was subtracted in order to compare with the solid line predictions from Sect.~\ref{sect:bb} ahead.

\subsection{Fluctuations}
\label{sec:fluctuations}

Consider now the fluctuations of ${\cal S}_i$.  The quantity that
actually sets the scale for the fluctuations in the residual
interaction energy is, as discussed in the introduction, approximately
the variance, ${\rm Var}[{\cal S}_i - {\cal S}_j]$.  It is understood
that $(i,j)$ do not differ by more than some small integer.  More
specifically, the interest is in computing this quantity to the
leading order in the semiclassical parameter $(\kf L)$. 

Using Eq.~(\ref{eq:sinr}) along with the decomposition into a secular
and fluctuating part of $N(\br,E_i^+)$ given by Eq.~(\ref{eq:nre2}),
results in ${\cal S}_i$ being written as the sum of two independent
terms.  The second one, ${\cal A}\int {\rm d}\br\ \left|\Psi_i(\br) \right|^2 \delta
N(\br;E_i^+)$, has been considered in \cite{Ullmo01prbb,Usaj02} within
the random plane wave approximation. It has a variance scaling as
$\log(\kf L)/(\kf L)$, which is of lower order than the leading behavior of ${\rm Var}[{\cal S}_i]$ calculated ahead.  Although, the second term is potentially of physical interest (see Sect.~\ref{sec:discussion}), the focus here is on developing the theory that gets the leading term analytically.  Applying
Eq.~(\ref{eq:psidef}) and dropping all the lower corrections gives for
the covariance between ${\cal S}_i$ and ${\cal S}_j$
\begin{equation}
\label{eq:var}
{\rm Covar}\left({\cal S}_i  {\cal S}_j\right) =  \frac{ ij}{ {\cal A}^2} \int {\rm
  d}\br_1\int {\rm d}\br_2\  \frac{J_1(2k_Fx_1) }{k_Fx_1}   \frac{J_1(2k_Fx_2)
  }{k_Fx_2}\left[ \left<\epsilon_i(\br_1) \epsilon_j(\br_2)\right>
  -\left<\epsilon_i(\br_1)\right>\left<
    \epsilon_j(\br_2)\right>\right] \; .
\end{equation}
As before, $x$ and $y$ are  the coordinates perpendicular and
parallel to the boundary.  Not surprisingly, since eigenstate-to-eigenstate 
correlations are not included in random plane wave modeling, 
\begin{equation}
\label{eq:epseps}
\left<\epsilon_i(\br_1) \epsilon_j(\br_2)\right>
-\left<\epsilon_i(\br_1)\right>\left< \epsilon_j(\br_2)\right> =
\delta_{ij} \left[ \left<\epsilon_i(\br_1) \epsilon_i(\br_2)\right>
  -\left<\epsilon_i(\br_1)\right>\left<
    \epsilon_i(\br_2)\right>\right] 
\end{equation}
and thus, ${\rm Var}[{\cal S}_i - {\cal S}_j]= 2{\rm Var}[{\cal S}_i]$.

Performing a little more algebra gives
\begin{equation}
\label{eq:epsilon}
\left<\epsilon_i(\br_1) \epsilon_i(\br_2)\right>
-\left<\epsilon_i(\br_1)\right>\left< \epsilon_i(\br_2)\right> = {\cal
  A}^2\left[\left<|\psi_i(\br_1)|^2 |\psi_i(\br_2)|^2\right>
  -\left<|\psi_i(\br_1)|^2\right>\left< |\psi_i(\br_2)|^2\right>
\right] 
\end{equation}
where Eq.~(\ref{eq:rpwm}) is applied to evaluate the right hand side 
of this equation.  Each resulting term has a product of four Gaussian 
random coefficients.  The fluctuations are thus given by pair-wise
correlating coefficients such that
\begin{eqnarray}
\label{eq:binarycor}
\langle a_l a_{l'} a_m a_{m'} \rangle &=& \langle a_l a_{l'} \rangle
\langle a_m a_{m'} \rangle + \langle a_l a_m \rangle \langle a_{l'}
a_{m'} \rangle + \langle a_l a_{m'} \rangle \langle a_{l'} a_m \rangle
\nonumber \\ 
&=& \sigma^4\left(\delta_{ll'} \delta_{mm'} + \delta_{lm}
  \delta_{l'm'} + \delta_{lm'} \delta_{l'm} \right) 
\end{eqnarray}
where $(l,l')$ are linked to the first coordinate, $\br_1$, and $(m,m')$ 
are linked to the second coordinate, $\br_2$.  The first term, which 
correlates the wavefunctions taken at the same position just
reproduces the mean $\left<|\psi_i(\br_1)|^2\right>\left<
  |\psi_i(\br_2)|^2\right>$ and cancels from Eq.~(\ref{eq:epsilon}).  The two
remaining terms give the same contribution, which can be understood
as a consequence of time reversal invariance.  Only one of those terms
would be non-zero for a time reversal non-invariant system, and
the result for the variance of ${\cal S}_i$ would just be
divided by two in that case.  

Therefore, together with 
averaging over $(\varphi_l, \varphi_2)$, the variance is 
\begin{eqnarray}
\label{eq:varianceexp}
{\rm Var}\left[{\cal S}_i \right] & = & \frac{8 i^2}{{\cal A}^2}
\frac{1}{N_{\rm eff}^2} \sum^{N_{\rm eff}}_{l , m=1} \int d\br_1 d\br_2
\frac{J_1(2\kf x_1)}{\kf x_1} \frac{J_1(2\kf x_2)}{\kf x_2}
\cs({\bf k}_l\cdot {\bf x}_1) \cs({\bf k}_l\cdot {\bf x}_2)  \nonumber \\
&&\qquad \times \cs({\bf k}_m\cdot {\bf x}_1) \cs({\bf k}_m\cdot {\bf
  x}_2)\cos\left[ {\bf k}_l \cdot ({\bf y}_1- {\bf y}_2)\right]
\cos\left[ {\bf k}_m \cdot ({\bf y}_1- {\bf y}_2)\right] \; . 
\end{eqnarray}
Reflection of either of the vectors $({\bf k}_l,{\bf k}_m)$ leaves the
integrand unchanged.  Thus, $\cos\left[ {\bf k}_l \cdot ({\bf y}_1-
  {\bf y}_2)\right] \cos\left[ {\bf k}_m \cdot ({\bf y}_1- {\bf
    y}_2)\right]$ is equivalent to $\exp\left[ i({\bf k}_l - {\bf
    k}_m)\cdot ({\bf y}_1- {\bf y}_2)\right]$ and can be replaced in
the integrand.  The summations can be replaced by angular integration
again as was done in Eq.~(\ref{eq:normalize}). 

For the purpose of understanding the asymptotic limit, one is tempted
to extend the limits of integration in these integrals. However, that
generates divergences associated with large $\delta y=y_1-y_2$ and
small $\sin(\theta_l-\theta_m)$ (where $\theta_{l,m}$ is the angle
between the vector ${\bf k}_{l,m}$ and the direction $\hat x$).  This
indicates that over large distances the random plane wave model as
given by Eq.~(\ref{eq:rpwm}) cannot be applied.  One way to think of
this is to imagine a true eigenstate of some chaotic billiard.
Locally, one could project onto the form of Eq.~(\ref{eq:rpwm}) and
approximately solve for a set of coefficients $\{a_l\}$ and plane wave
orientations.  However, the solution set $\{a_l\}$ would be dependent
on the location along the boundary where the projection was performed
due to the rotating orientation of the local coordinate system.  Even
if the Gaussian random modeling were perfectly fine from
state-to-state, as $\br_2$ got further from $\br_1$, the two
cross-terms in Eq.~(\ref{eq:binarycor}) that generate the variance
would progressively decay on a length scale given by the typical
dimension $L$ of the system.  This is related to the behavior observed
for the spatial autocorrelation function; see Fig.~2 of \cite{BaeSch2002b}.
(Note that the first term, which reproduces the square of the mean
would on the other hand not decay.)  Ahead, it is seen that the
results depend only logarithmically on this parameter for the Neumann
boundary conditions and not at all for Dirichlet, so that it is not
necessary to describe very precisely this decay as long as the proper
length scale is introduced.

A Gaussian form 
$\exp(-\delta y^2/2L^2)$ is convenient and gives 
\begin{eqnarray}
\label{eq:varianceexp2}
{\rm Var}\left[{\cal S}_i \right] & = & \frac{8 i^2}{\pi^2{\cal A}^2}
\int_{-\pi/2}^{\pi/2}d\theta_l \int_{-\pi/2}^{\pi/2}d\theta_m \left[
  \int_0^\infty dx  \frac{J_1(2\kf x)}{\kf x}  \cs(\kf x\cos\theta_l)
  \cs(\kf x\cos\theta_m) \right]^2  \nonumber \\ 
&&\qquad \times {\cal L} \int^{{\cal L}/2}_{-{\cal L}/2} d(\delta y)
\exp\left(-\frac{\delta y^2 }{ 2L^2} + i({\bf k}_l - {\bf k}_m)\cdot
  ({\bf y}_1- {\bf y}_2) \right) \; . 
\end{eqnarray}
Including the Gaussian cutoff, and noting that the dominant contributions come from regions in which $\sin\delta\theta$ is small (with $\delta \theta = \theta_l - \theta_m$),
it is possible to approximate $\sin\theta_l-\sin\theta_m \simeq \cos(\bar \theta) \delta\theta$ with $\bar \theta = (\theta_l+ \theta_m)/2$.  The integrand $\cal I(\bar\theta,\delta\theta)$ becomes
\begin{equation} \label{eq:Int2bis}
{\cal I}(\bar\theta,\delta\theta) = \sqrt{2\pi}L{\cal L} \left[\frac{1 \pm
  |\sin(\bar\theta)|}{2\kf}\right]^2 
    \exp\left[-\frac{\left(\kf L \cos\bar\theta\delta\theta\right)^2 }{2} \right]
\end{equation}
where the sign $-$ and $+$ correspond to Dirichlet and Neumann
boundary conditions respectively.  Performing the integration over the
variables $\delta \theta$ and $\bar \theta$ yields
\begin{equation}
{\rm Var}[{\cal S}_i]  = 
\frac{\kf {\cal L} }{4\pi^3} \langle \lambda^2(\bar \theta) \rangle_\theta
\label{eq:Var} 
\end{equation}
where we have introduced the function
\begin{equation} \label{eq:lambda}
\lambda (\theta) \df \frac{\left[{1 \pm
      |\sin(\theta)|}\right]}{|\cos(\theta)|} 
\end{equation}
 and the average is defined in terms of the variable  $\sin(\theta)$ so that
\begin{eqnarray}
\langle \lambda^2(\theta) \rangle_\theta = \int_0^1 d(\sin \theta)
\lambda^2(\theta)  
 & = & (2\ln 2 - 1)
\qquad \qquad \qquad \quad\  \mbox{Dirichlet}  \label{eq:VarDir}\\
& = & (2\ln 2 - 1)  + 4\ln \frac{\pi \kf {\cal A}}{ 2{\cal L}}
   \qquad \mbox{Neumann}  \label{eq:VarNeu} \; .
\end{eqnarray}
Note caution must be exercised in evaluating the Neumann case.  
There the angle $\bar\theta$ cannot be allowed to decrease to less 
than the inverse of $\kf L$ where the cutoff expression becomes
invalid.  In the absence of other considerations, a very reasonable
choice for $L$ is just half the average length between two reflections; 
see Appendix~\ref{app:meanl}.  For a 2D concave
billiard, this gives exactly $L=\pi{\cal A} /(2{\cal L})$, and this
value has been substituted into the Neumann form. 

Finally, consider the difference between Dirichlet and
Neumann boundary conditions.  Equations~(\ref{eq:VarDir},\ref{eq:VarNeu}) both roughly imply a $\kf {\cal L}$ behavior.  The
prefactor $\ln (2/\sqrt{e})/2\pi^3 \simeq 0.0031$ is rather small in
the first case, whereas for Neumann boundary conditions there is a
logarithmic enhancement which can be understood as a (much larger)
effective prefactor (a factor 40 larger for $i=1000$).  From the point of view of the calculation, the
difference between these two cases can be related to the sign change
between $1- |\sin(\bar\theta)|$ and $1+ |\sin(\bar\theta)|$ in
Eq.~(\ref{eq:Int2bis}), in such a way that whispering gallery modes
(for which the corresponding classical orbits have $\bar\theta \simeq \pi/2$) 
are suppressed for Dirichlet boundary
conditions whereas they dominate (because they are less affected by
the $\exp[-(\kf L \cos\bar\theta\delta\theta)^2/2]$ factor) in the
Neumann case.  This makes sense since the main source of ${\cal S}_i$
fluctuations originates from the wavefunction fluctuations' probability
density $|\psi_i(\br)|^2$ in the mean field $ \propto[(1 \pm
J_1(2\kf x)) / (\kf x)]$ generated by the Friedel oscillations of all
the other particles below the Fermi energy.  Dirichlet boundary
conditions however impose that $|\psi_i(\br)|^2 \to 0$ as $\br$
approaches the boundary and therefore inhibits this contribution.

\begin{figure}[t]
\PSImagx{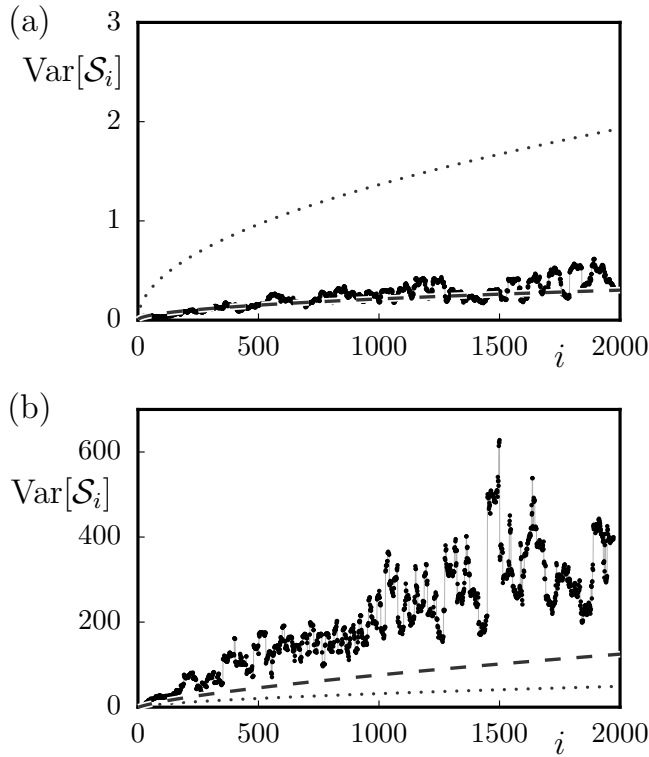}{8.6cm}
\caption{Variance of ${\cal S}_i$ for the cardioid (odd parity only) and quarter stadium (even-even symmetry only) billiards using Dirichlet boundary conditions. In (a), the discrete points are the cardioid billiard results, the dotted line is the result of the random plane wave model, i.e. Eq.~(\ref{eq:Var}), and the long dashed line is the semiclassical theory, Eq.~(\ref{eq:variancesemi},\ref{eq:1/2classics}), given in Sect.~\ref{cqb} ahead. In (b), the discrete points are the stadium billiard results, the dotted line is the result of the random plane wave model, Eq.~(\ref{eq:Var}), and the long dashed line is the prediction Eq.~(\ref{eq:variancefb}) from the semi-quantitative semiclassical theory developed in Sect.~\ref{sect:bb} ahead for the bouncing ball modes. }
\label{fig:varianceSi}
\end{figure}

Figure \ref{fig:varianceSi} illustrates the comparison between the analytical
results of the random plane wave model for the variance, Eq.~(\ref{eq:Var}), for the cardioid and quarter stadium billiards.  For the stadium, $\kf {\cal L}$ is replaced by $\kf {\cal L}_N$, i.e.~the length of the straight edges where Neumann boundary conditions are imposed (even-even symmetry class).  The theory for the Dirichlet case, cardioid billiard, appears to be roughly a factor six too great.  In order to understand the discrepancy, the more powerful approach of semiclassical theory is developed in the next section.  For the stadium, both Neumann boundary conditions and bouncing ball modes must be considered.  This involves additional complications treated in Sect.~\ref{sect:bb} ahead.    

\section{Semiclassical approach}
\label{sec:gutzwiller}

It is important to develop a semiclassical approach.  It gives a more powerful theory and sheds some light on the difficulties that the random plane wave model is having in providing a
quantitative description of the $\Si$ fluctuations.  The most
immediate conceptual difficulty in getting started is that a treatment
of the $\Si$ implies, through Eqs.~(\ref{eq:psidef},\ref{eq:Si}),
addressing the fluctuations of {\em individual wavefunctions}, whereas
semiclassical approximations valid for chaotic systems -- such as the
ones based on the semiclassical Green functions -- converge only for
quantities smoothed on an energy range containing a significant number
of levels. Here, however, this difficulty can be overcome.  

\begin{figure}[t]
\PSImagx{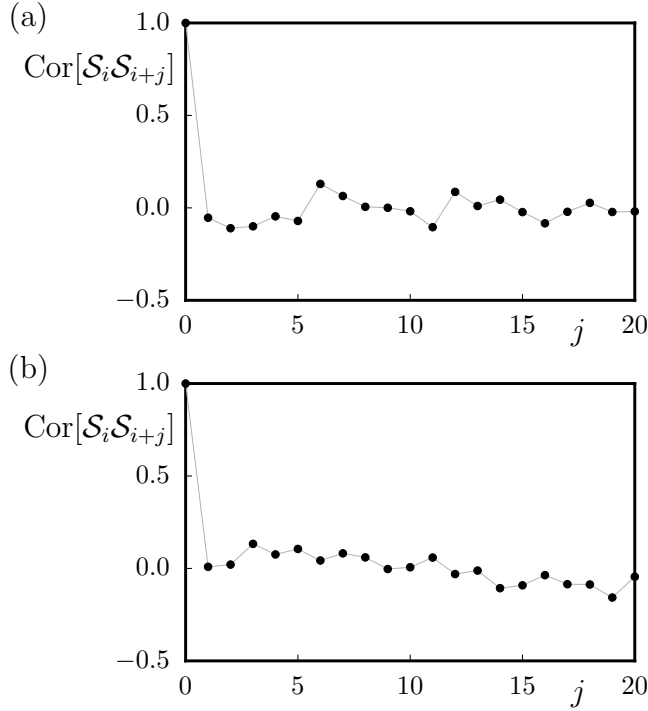}{8.6cm}
\caption{Correlation function ${\rm Cor}\left[ {\cal S}_i {\cal S}_{i+j}\right] = {\rm Cov}\left[ {\cal S}_i {\cal S}_{i+j}\right]/{\rm Var}\left[ {\cal S}_i\right]$ for the cardioid and stadium billiard examples.  In both cases the correlation function (or covariance) of the billiard is consistent with zero to within sample size fluctuations as assumed in the random plane wave model, and as implied by the semiclassical theory ahead leading to Eq.~(\ref{eq:varSN2}). }
\label{fig:covar}
\end{figure}

For this purpose, let us, following Bogomolny \cite{Bogomolny88}, introduce a local energy averaging that is generally much narrower than the Thouless energy
\begin{equation}
\label{eq:smean}
\overline{\cal S}_{\Delta N} \df \frac{1}{\Delta N} \sum_{E
  -\frac{\Delta E}{2} < E_i < E +   \frac{\Delta E}{2}}   \Si \; 
\end{equation}
The notation $\Delta N \df N(E+ \frac{\Delta E}{2}) - N(E-
\frac{\Delta E}{2})$ represents the number of levels in the energy
interval $[E -\frac{\Delta E}{2}, E + \frac{\Delta E}{2}]$.  With this notation, the variance is
\begin{equation} \label{eq:transmuttation}
 {\rm Var} [\ \overline{\cal S}_{\Delta N}] \df 
\left\langle \left (\ \overline{\cal S}_{\Delta N}  - \langle {\cal S} \rangle \right)^2 \right\rangle
= \frac{1}{\Delta N} \varSi + \frac{\Delta N - 1}{\Delta N}
{\rm Covar}_{i \neq j} [{\cal S}_i {\cal S}_j] \; 
\end{equation}
($\langle {\cal S} \rangle \df \langle {\cal S}_i \rangle = \left\langle \ \overline{\cal S}_{\Delta N} \right\rangle $).
Computing the locally averaged quantity $\overline{\cal S}_{\Delta N}$, for which convergent
semiclassical approximations can be used, it is possible to extract
the variance and the covariance of the $\Si$ from the scaling in
$\Delta N$ of $ {\rm Var} [\ \overline{\cal S}_{\Delta N}] $.  In addition, as expected from the random plane wave description, the correlations amongst the $\Si$ are entirely negligible. This is
illustrated in Fig.~\ref{fig:covar},
where the average for the covariance is performed over 400 levels starting
from $i=1500$. Both cardioid and stadium billiard exihibit 
correlations which are within the expected statistical errors for being
consistent with zero. It is 
therefore expected (and actually turns out) that the semiclassical
evaluation of $ {\rm Var} [\ \overline{\cal S}_{\Delta N}]$ 
scales as $1/\Delta N$, and it is possible to interpret the corresponding
multiplicative factor as $\varSi$.

The remaining task is to evaluate semiclassically the locally
smoothed quantity $\overline{\cal S}_{\Delta N}$.  For this purpose, two ingredients are
needed, the wavefunction probabilities $|\Psi_i(\br)|^2$ and the
density of particles $N(\br,E)$.  For the eigenfunctions, a
semiclassical orbit summation was given by Bogomolny
\cite{Bogomolny88}.  It is based on the semiclassical approximation of
the retarded Green's function (given here for two dimensional systems)
\begin{equation} \label{eq:G1/2classic}
G^R(\br,\br',E) \simeq 
\frac{1}{i\hbar} \frac{1}{\sqrt{2i\pi\hbar}  }
\sum_{\mu: \br \to \br'}  \frac{1}{ \sqrt{|\dot x \dot x' m_{12,\mu}|}} \exp\left[
  \frac{i}{\hbar} S_\mu(\br,\br',E) - i \frac{\pi}{2} \eta_\mu\right ]
\end{equation}
where the sum runs over all {\em closed} (i.e. not necessarily
periodic) orbits, with $S_\mu$ the classical action of the $\mu^{th}$
orbit, $m_{12,\mu} \equiv \partial r_\perp'/\partial p_\perp$ the
stability matrix element, and $\eta_\mu$ the appropriate geometric index
(primed and unprimed variables correspond respectively to the initial
and final coordinates).   $G^R(\br,\br,E)$ is related to the local density
of states, and thus to the eigenfunction probability density via
\begin{equation} \label{eq:nu1/2classic} 
  \nu(\br,E) \equiv \sum_i
  |\Psi_i(\br)|^2 \delta (E-E_i) =- \frac{1}{\pi} {\rm Im}\
  G^R(\br,\br,E) \; . 
\end{equation}
Introducing the density of states $\rho(E) \equiv \sum_j
 \delta (E-E_j)$, we obtain
\begin{equation} \label{eq:wf_def} 
  \overline{|\Psi_i(\br)|^2}_{\Delta N}  = 
  \frac{\overline{\nu(\br,E_i^+)}_{\Delta E} }{\overline{\rho(E_i^+)}_{\Delta E} }, 
\end{equation}
where on the r.h.s. the overline notation has the same meaning as
previously introduced except that the division is by $\Delta E$
instead of $\Delta N$.  

In the semiclassical evaluation of $\nu(\br,E)$, it is typical to
distinguish between the ``zero-length'' orbit contribution $\nu_W(\br)
= m/2\pi\hbar^2$ and the contribution of the remaining orbits, whose
lengths remain finite as $\br \to \br'$.  Here however, 
interest is in the fluctuations of the eigenfunctions, and thus of
$\nu(\br,E)$, near the boundary of the billiard.  The orbit
responsible for the Friedel oscillations, namely the one returning to
its initial location immediately after bouncing off the boundary will
therefore be extremely short, implying that: i) the corresponding
contribution will not be sensitive to local energy averaging ; and ii)
if $\br$ is at a distance $x$ from the boundary shorter than, or of
the order of, the Fermi wavelength, the semiclassical approximation
Eq.~(\ref{eq:G1/2classic}) cannot be applied.  On the other hand,
assuming $x$ much smaller than the curvature of the boundary, this contribution can be approximated by the exact result valid (in two dimension) for a straight boundary $\nu_{\rm Friedel}(x) = \pm \nu_W
J_0(2\kf x)$.  This gives
\begin{equation} \label{eq:nu(r)}
\overline{\nu(\br)}_{\Delta E}  = \nu_W \left[1 \pm J_0(2\kf x)
\right] - \frac{1}{\pi} {\rm Im}\ \overline{\tilde G_{\rm osc}(\br,\br)}_{\Delta E} \; ,
\end{equation}
valid near the billiard boundary.  The tilde on $\tilde G_{\rm osc}$ indicates that the short orbits giving rise to Friedel oscillations have been excluded from the semiclassical sum Eq.~(\ref{eq:G1/2classic}).  

Similarly, the density of states is split into a smooth component and an oscillatory one.  Once the local energy averaging is performed over a range much larger than the mean level spacing,  the oscillatory components are small compared with the smooth term.  Thus, the density of states can be expanded in the denominator.    Addressing still two-dimensional billiard systems for which $\rho_{\rm W}(E) = { \cal
  A} \nu_W$ gives 
\begin{equation}
\label{eq:wf1} 
{\cal A}  \overline{|\Psi_i(\br)|^2}_{\Delta N}
= 1 \pm J_0(2\kf  x) - \frac{1}{\nu_W } \frac{1}{\pi} {\rm Im}\
\overline{\tilde G_{\rm  
  osc} (\br,\br,E)}_{\Delta E}  - \frac{1 \pm J_0(2\kf
  x)}{{\cal A} \nu_W} \overline{\rho_{\rm osc}(E)}_{\Delta E} \; .
\end{equation} 
This equation could be thought of as a slight generalization of the
result given by Bogomolny~\cite{Bogomolny88}, with the only difference
that the Bessel function $J_0(2\kf x)$ has been introduced to account
for the Friedel oscillations (which turn out to be important here); see Appendix~\ref{app:norm} for an improved normalization of this equation.

For $\Delta E$ large on the scale of the mean level spacing, but small
on the classical scale, the energy smoothing can be performed for each orbit
contribution noting that $\partial S_\mu/\partial E = \tau_\mu$, with $\tau_\mu$
the time of travel of the orbit, giving
\begin{equation} \label{eq:damping}
\overline{\exp \left( \frac{i}{\hbar}  S_\mu \right)}_{\Delta E} = \exp \left( \frac{i}{\hbar}
\bar S_\mu \right)  {\rm sinc}\left(\frac{\tau_\mu   \Delta E}{2 \hbar} \right) \; ,
\end{equation}
with [${\rm sinc}(x) \df \sin(x)/x$].  Energy smoothing therefore
implies that orbits with periods greater than $\hbar/\Delta E$ are cut
off in the semiclassical sums
Eqs.~(\ref{eq:G1/2classic}-\ref{eq:wf1}).

As a direct (and expected) consequence, if the smoothing takes place
on a energy range larger than the Thouless energy, no orbit can
contribute to $\overline{\tilde G_{\rm osc} (\br,\br,E)}_{\Delta E}$
or $\overline{\rho_{\rm osc}(E)}_{\Delta E}$, and the average
wavefunction probability reduces to ${\cal A} \langle |\Psi_i(\br)|^2
\rangle = 1 \pm J_0(2\kf x)$.  Inserting this equality into
Eq.~(\ref{eq:Si}) with the definition Eq.~(\ref{eq:psidef}), we
readily obtain the result Eq.~(\ref{eq:simean}) for the mean value
$\left<{\cal S}_i \right>$, 
but here without any assumption regarding the nature of the dynamics; i.e.
it applies equally well for integrable, mixed or chaotic systems.

\subsection{Chaotic quantized billiards}
\label{cqb}

Up to this point, the nature of the dynamics has played no role in the semiclassical approach.  However, beginning here, the approach is specialized to chaotic systems.  The oscillating component (a sum over periodic orbits~\cite{Gutzwiller90}) of the density of states is given by
\begin{equation} 
\label{eq:rho_osc} 
\rho_{\rm osc}(E) =  \frac{1}{\pi\hbar}  \sum_{\gamma={\rm
    periodic \, orbit}} \frac{\tau_\gamma}{ 
 \left| \Det\left(M_\gamma - {\bf 1}\right)\right|^{1/2}} \cos\left(
 \frac{S_\gamma(E^+_i)}{\hbar} - \eta_\gamma \frac{\pi}{ 2}\right) 
\end{equation}
where $\tau_\gamma$ is the period of the periodic orbit, $M_\gamma$ the monodromy
matrix and $\eta_\gamma $ the appropriate geometric index.  

For a two-degree-of-freedom billiard, ${\cal A}
\overline{\left|\Psi_i(\br)\right|^2}_\Nsm 
\simeq 1 \pm J_0(2\kf x) + \overline{\epsilon^{(1)}_i(\br)}_\Nsm +
\overline{\epsilon^{(2)}_i(\br)}_\Nsm $ with
\begin{eqnarray}
   \overline{\epsilon^{(1)}_i(\br)}_\Nsm &\approx&
   \frac{2\sqrt{\hbar}}{ m}    {\rm Im} \frac{i }{ \sqrt{2\pi i}} \sum_{\mu = {\rm closed \, orbit}} \frac{1}{
    \sqrt{|\dot x_\mu \dot x'_\mu m_{12,\mu}|}} \overline{\exp\left[i \frac{S_\mu(\br,\br)}{
      \hbar}-i\nu_\mu\frac{\pi}{2}  \right]}_{\Delta E}  \\   
\overline{\epsilon^{(2)}_i(\br)}_\Nsm  &\approx&
\frac{2 \hbar}{m {\cal A}}  \left(1 \pm J_0(2\kf x) \right)
     \sum_{\gamma = {\rm periodic \, orbit}} \frac{\tau_\gamma}{ 
 \left| \Det\left(M_\gamma - {\bf 1}\right)\right|^{1/2}} \overline{\cos\left(
 \frac{S_\gamma}{\hbar} - \bar\nu_\gamma
 \frac{\pi}{2}\right)}_{\Delta E}  \; . \nonumber \\
\label{eq:epsi}
\end{eqnarray}
Note that the $\mu$-orbit sum here includes all returning orbits, not
just those in the neighborhood of a complete periodic orbit.

\begin{figure}[t]
\PSImagx{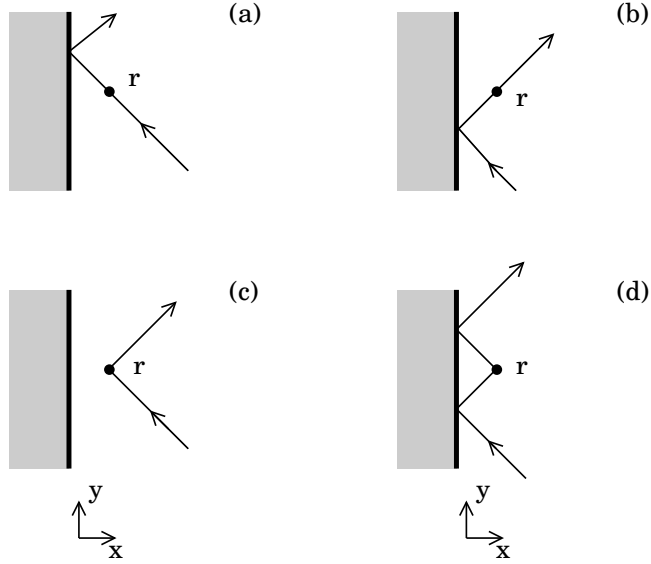}{8.6cm}
\caption{Sketch of the four orbits which, as $x \to 0$, coalesce into
  the same [nearly] periodic orbit. The top row corresponds to two
  (nearly) periodic orbits such that $\br$ lies on the trajectory
  either (a) just before or (b) just after the bounce off the
  boundary. The bottom row corresponds to two non-periodic orbits
  ($p'_x =- p_x)$ such near $\br$ (c) one of them does not touch the boundary
  and (d) the other one touches twice.}
\label{fig:orb_clust}
\end{figure}

One might also be tempted to use the same expressions integrated over
energy to deduce a similar expression for $N(\br;E^+_i)$. However, the
energy integral generates a factor $\sim \hbar/\tau_i E$ for the
oscillating contribution of an orbit of period $\tau_i$, and therefore
the oscillating terms obtained in this way would be of lower order in
$\hbar$ than those generated by $|\Psi_i(\br)|^2$. In leading order,
it is therefore only necessary to keep the terms of $N(\br;E^+_i)$ 
associated with the Friedel oscillations, i.e.~begin with
Eq.~(\ref{eq:Si}) directly and drop further sub-leading terms.  Thus, ${\cal S}_i = \overline{\cal S} \pm \left[  {\cal S}^{(1)}_{i,{\rm osc}}  + {\cal S}^{(2)}_{i,{\rm osc}}  \right]$,
where
\begin{equation} \label{eq:Si_simple}
{\cal S}^{(\alpha)}_{i,{\rm osc}} = 
\frac{i}{\cal A}\int {\rm d}\br \,  \frac{J_1(2 \kf x)}{(\kf x)} \,
\epsilon^{(\alpha)}_i(\br) \; ,
\end{equation}
($\alpha=1,2$).  Here, two remarks are in order.  First, because only
the very short orbit contribution (i.e the term proportional to $J_1(2
\kf x)/(\kf x)$) is kept for $N(\br,E)$, it is not sensitive
to the local energy average and can be taken out of the bracket.
Secondly, note that the main contribution to the integral over space in
the r.h.s.  of Eq.~(\ref{eq:Si_simple}) is restricted to the vicinity
of the boundary.  As before we can unambiguously use a system
of coordinates $\br=(x,y)$ with $x$ perpendicular and $y$ parallel to
the boundary.  To compute ${\cal S}^{(1)}_{i,{\rm osc}}$, insert
Eq.~(\ref{eq:epsi}) into Eq.~(\ref{eq:Si_simple}).  A stationary phase
condition has to be imposed in the $y$ direction, but not in the $x$
direction since the effective range of interaction is not large,
even on the scale of the Fermi wavelength.  As a consequence, the
dominant contributions of the integration involved in
Eq.~(\ref{eq:Si_simple}) come from the neighborhood of trajectories
such that $p'_y = p_y$ (where the primed (unprimed) correspond to
initial (final) momentum), but for which the initial and final
$x$-momenta may differ.  Energy conservation however imposes $p'_x =
\pm p_x$.  As illustrated in Fig.~\ref{fig:orb_clust}, these
trajectories can be associated to a cluster of four orbits which, as $x
\to 0$, converges smoothly toward the same nearly periodic orbit (or
fixed point of the boundary Poincar\'e map): i) two nearly periodic
orbits such that $\br$ lies on the trajectory just before or just
after bouncing off the boundary, and ii) two non-periodic ones (
$p'_x =- p_x)$ either touching the boundary twice or not at all near $\br$.

Denoting $S_{l_0}(0,y)$ the action of the nearly periodic orbit to which all of
these orbits converge as $x \to 0$, leads to $S_l(x,y) =
S_{l_0}(0,y) + \delta S_l(x,y)$, where $ \delta S_l \simeq (p'_x -
p_x) x$, which vanishes for the two periodic orbits and gives $\pm 2 |p| x
\cos \theta_l $, with $\theta_l$ the angle of incidence of the periodic
orbit on the boundary, for the two non-periodic ones.  Noting that
$\exp (i 2 k x \cos \theta_l) + \exp (-i 2 k x \cos \theta_l) \pm 2 = 4 {\rm
  cs}^2( k x \cos \theta_l)^2$ gives
\begin{eqnarray}
\label{eq:gutz}
\overline{ {\cal S}^{(1)}_{i,{\rm osc}}}_\Nsm
  &=& \frac{8\sqrt{\hbar} }{m {\cal A}} \sum_{l= {\rm fixed \, point}} 
{\rm  sinc}\left(\frac{\tau_l   \Delta E}{2 \hbar} \right) 
 \int_0^\infty dx \frac{J_1(2kx)}{kx} \cs^2 (kx\cos\theta_l)
  \nonumber \\ 
&&\times  {\rm Im} \frac{i }{  \sqrt{2\pi i}} \int^{{\cal L}/2}_{-{\cal
    L}/2} dy  \frac{1 }{   \sqrt{|\dot{x}_l||\dot{x}'_l||m_{12,l}|}}
\exp\left[i \frac{S_l((0,y),(0,y);E) }{
    \hbar}-i\nu_l\frac{\pi}{ 2}\right] \,
\end{eqnarray} 
where the sum runs over all the fixed points of the boundary Poincar\'e
section.  As in the previous section $\int_0^\infty dx\frac{J_1(2kx)}{kx}  \cs^2 (kx\cos\theta_l)  = 
[1 \pm |\sin \theta_l|]/2k$.  Furthermore, the integral in the
parallel direction can be performed in a very similar way as in the
derivation of the Gutzwiller trace formula.  Using the fact that near the
periodic point $(0,y_l)$
\begin{equation}
\label{eq:action}
S_l(((0,y_l + \delta y),(0,y_l + \delta y);E) = S_l(E) + \frac{\Det(M_l -
  1)  }{2 m_{12,l}} \delta y^2
\end{equation}
and $|\dot x_l| = |\dot x'_l| = v_{\rm F} |\cos \theta_l|$
we get 
\begin{equation}
\label{eq:gutz2}
\overline{ {\cal S}^{(1)}_{i,{\rm osc}}}_\Nsm = 
\frac{4i }{{\cal A} \kf^2} \sum_{l = {\rm fixed \, point }} 
\lambda(\theta_l)
\frac{\cos\left[\frac{S_l(E)}{\hbar} - 
\bar\nu_l \frac{\pi}{ 2}\right]}{\sqrt{|\Det(M_l-1)|}} 
{\rm  sinc}\left(\frac{\tau_l  \Delta E}{2 \hbar} \right)  \; ,
\end{equation}
with $\lambda(\theta) \df \left[{1 \pm |\sin \theta|}\right] /{|\cos
  \theta|} $ the same function that was introduced in the random plane
wave approach (cf. Eq.~(\ref{eq:lambda})).

The computation of $\overline{ {\cal S}^{(2)}_{i,{\rm osc}}}_\Nsm$ is even simpler as
the only spatial dependence of $\epsilon_i^{(2)}$ arises from the
Bessel function $J_0(2\kf x)$.  To facilitate the comparison with
Eq.~(\ref{eq:gutz2}), replace the sum over periodic orbits by a
sum over fixed points of the Poincar\'e section, in which case the
period $\tau_\gamma$ of the periodic orbit has to be replaced by the
average time of flight $\tau_\gamma / n_\gamma = \ell_\gamma/v_{\rm F}
n_\gamma$ ($\ell_\gamma$ and $n_\gamma$ are respectively the total
length and total number of bounces of the peridic orbit $\gamma$)
between two successive bounces on the boundary.  Using $\int_0^\infty ({du}/{u}) \, J_1(u)(1 \pm J_0(u)) = (1 \pm 2/\pi)$ gives
\begin{equation}
\label{eq:gutz3}
\overline{ {\cal S}^{(2)}_{i,{\rm osc}}}_\Nsm  = -
\frac{4i }{{\cal A} \kf^2} \sum_{l = {\rm fixed \, point }} 
 \frac{{\cal L} \ell_l}{2 n_l {\cal A}} \left(1 \pm \frac{2}{\pi}\right)
\frac{\cos\left[\frac{S_l(E)}{\hbar} - 
\bar\nu_l \frac{\pi}{ 2}\right]}{\sqrt{|\Det(M_l-1)|}} 
{\rm  sinc}\left(\frac{\tau_l   \Delta E}{2 \hbar} \right) 
 \; ,
\end{equation}
where $n_l \df n_{\gamma(l)}$ is the number of bounces of the periodic
orbit to which $l$ belongs, and a similar notation is implied for the
other parameters.   This actually depends only on the periodic orbit
$\gamma$ and not on the specific periodic point $l$ on $\gamma$. The
$(2/\pi)$ factor can be traced back to the $J_0(2\kf x)$ term, and
therefore eventually to the Friedel oscillations of the local density
of states.

The two terms can be combined to give
\begin{eqnarray}
\overline{\cal S}_{\Delta N} - \langle {\cal S} \rangle =  \pm
\frac{4i}{{\cal A} \kf^2}  
&& \sum_{\gamma = {\rm periodic \, orbit}}
\frac{\cos\left[\frac{ S_\gamma(E)}{\hbar} -  \bar\nu_\gamma
      \frac{\pi}{ 2}\right]}{\sqrt{|\Det(M_\gamma-1)|}}  \,
 {\rm  sinc}\left(\frac{\tau_\gamma   \Delta E}{2 \hbar} \right)
\nonumber \\
& \times &  
\sum_{l =  {\rm fixed \, point \, of \,} \gamma}    \left[
  \lambda(\theta_l) - \frac{{\cal L} \ell_l }{ 
    2{\cal A} n_l}\left( 1\pm {2\over \pi} \right)  \right]
\; .
\label{eq:semiresult}
\end{eqnarray}
This form is suitable for performing the calculation of the variance.  However, as rederived briefly in Appendix~\ref{app:meanl}, note that the mean length per bounce in a billiard is 
\begin{equation} \label{eq:meanl}
\bar d=\frac{\pi {\cal A}}{\cal L} \; .
\end{equation}
Also, the density of fixed points is uniform for long
orbits in the measure ${\rm d}\sin\theta$.  Averaging with that
measure while  replacing ${{\cal L} \ell_l }/{
    2{\cal A} n_l}$ by its mean value $\pi/2$ gives
\begin{eqnarray}
\left\langle \lambda(\theta_l) - {{\cal
      L} \ell_l \over 2{\cal A} n_l}\left( 1\pm {2\over \pi} \right)
\right\rangle_\theta &=& \frac{1}{ 2} \int_{-\pi/2}^{\pi/2} {\rm d}\sin\theta
\frac{(1 \pm |\sin \theta_l|)}{|\cos \theta_l|}  -{\pi \over 2}\left(
  1\pm {2\over \pi} \right)  
\nonumber \\ 
&=& {\pi \over 2} \pm 1 -{\pi \over 2}\left( 1\pm {2\over \pi} \right)
= 0
\label{eq:savesourday} 
 \end{eqnarray}
 Thus, the constant $\frac{{\cal L}\bar d }{2{\cal A}}\left(1\pm \frac{2}{\pi} \right)$ can be understood as arising from $\overline{ {\cal S}^{(2)}_{i,{\rm osc}}}_{\Delta E}$ (i.e.~associated with the density of states)  as the angular mean $\langle \lambda(\theta) \rangle_\theta$ of the corresponding term in $\overline{ {\cal S}^{(1)}_{i,{\rm osc}}}_{\Delta E}$.

To compute the variance, it is necessary to square Eq.~(\ref{eq:semiresult}) and average the resulting expression over a large energy range.  For two periodic orbits $\gamma$ and $\gamma'$, 
 $\left\langle {\cos\left[\frac{S_\gamma(E)}{\hbar} - \bar\nu_\gamma \frac{\pi}{
         2}\right]} {\cos\left[\frac{S_{\gamma'}(E)}{\hbar} - \bar\nu_{\gamma'}
       \frac{\pi}{ 2}\right]}\right\rangle$ equals one half if $\gamma$ and
 $\gamma'$ are either the same orbit or time reversal symmetric, but
 zero otherwise, which
 makes cross-terms from different periodic orbits vanish.  Note
 however that it does not eliminate cross-terms of the various fixed
 points for a given orbit since these contributions oscillate with
 the same frequency.   This gives
\begin{eqnarray} 
\left\langle \left(\ \overline{\cal S}_{\Delta N} - 
\langle {\cal S} \rangle \right)^2 \right\rangle & = & \frac{g_s}{2}
\frac{16 i^2}{k^4A^2}
\sum_{{\rm orbit} \gamma} \frac{1} {|\Det(M_l-1)|} 
{\rm  sinc}^2\left(\frac{\tau_\gamma \Delta E}{2 \hbar} \right) \nonumber \\ 
& \times & 
\sum_{l,l' \atop \, {\rm fixed \,  point \,  of \, } \gamma} 
  \left[\lambda(\theta_l) - \frac{{\cal L} \ell_\gamma }{2{\cal A}
      n_\gamma}\left( 1\pm {2\over \pi} \right) \right]
\left[\lambda(\theta_{l'}) - \frac{{\cal L} \ell_\gamma }{2{\cal A}
      n_\gamma}\left( 1\pm {2\over \pi} \right) \right] \; .  \nonumber \\  \label{eq:varS1}
\end{eqnarray}
For long orbits, which are going to dominate this sum, it is possible to identify $\ell_\gamma / n_\gamma$ with $\bar d=\pi {\cal A}/{\cal L}$, the mean length per bounce in the billiard, and assume
that the angles $\theta_l$ are uncorrelated and uniformly distributed
with the measure ${\rm d}\sin\theta$.  It turns out for the
last sum in Eq.~(\ref{eq:varS1}) 
\begin{eqnarray}
 \sum_{l,l' \atop \, {\rm fixed \,  point \,  of \, } \gamma} &&
  \left[\lambda(\theta_l) - \frac{{\cal L} d_\gamma }{2{\cal A}
      n_\gamma}\left( 1\pm {2\over \pi} \right) \right]
\left[\lambda(\theta_{l'}) - \frac{{\cal L} d_\gamma }{2{\cal A}
      n_\gamma}\left( 1\pm {2\over \pi} \right) \right] \nonumber \\
& \simeq & 
n_\gamma \left\langle \left[\lambda(\theta) - \frac{{\cal L} \bar d
  }{2{\cal A}}\left( 1\pm {2\over \pi} \right) \right]^2 \right\rangle_\theta + 
n_\gamma (n_\gamma - 1) 
\left\langle \left[\lambda(\theta) - \frac{{\cal L} \bar d
  }{2{\cal A}}\left( 1\pm {2\over \pi} \right) \right] \right\rangle_\theta^2  \nonumber \\
& =  &
n_\gamma \left\langle \left(\lambda(\theta) -  \langle
  \lambda\right\rangle_\theta  \right)^2 \rangle_\theta   \; ,
\label{eq:nocrosscor}
\end{eqnarray}
where Eq.~(\ref{eq:savesourday}) has been used to cancel the cross
terms between different fixed points.  Making use of the Hannay-Ozorio de Almeida sum rule~\cite{OzorioBook} in the form
\begin{equation} \label{eq:HannayOzorio}
\sum_{{\rm fixed \, points} \, l \atop {\rm with} \, n \; {\rm
    bounces}} \frac{1} {|\Det(M_l-1)|} = 1 \; ,
\end{equation}
(where the sum runs over all fixed points belonging to a periodic
orbit with $n$ bounces), identifying the period $\tau$ of the orbit
with $n \bar d / \vf$ ($\vf = \hbar \kf / m$ is the Fermi velocity),
replacing the sum over the number of bounces by an integral, and
making use of Eq.~(\ref{eq:meanl}) gives
\begin{eqnarray} 
\overline{\left( \langle {\cal S} \rangle_{\Delta N} - \overline{\cal
      S}  \right)^2} & = & \frac{g_s }{2}
\frac{16 i^2}{k^4A^2}  \left\langle \left(\lambda(\theta) -  \langle
  \lambda\right\rangle_\theta  \right)^2 \rangle_\theta  
\int dn \, {\rm sinc}^2 \left(\frac{ n \bar d \Delta E}{2 \hbar \vf
  } \right)  \label{eq:varSN1} \nonumber \\
 & = & \frac{g_s }{2}
{\kf {\cal L} \over 2 \pi^3} 
\left\langle \left(\lambda(\theta) -  \langle
  \lambda \rangle_\theta  \right)^2 \right \rangle_\theta \,
\frac{1}{\Delta N} \; .  \label{eq:varSN2}
\end{eqnarray}
As expected, the variance of $\overline{\cal S}_{\Delta N}$ is
inversely proportional to $\Delta N = \rho_W(E) \Delta E$.  Applying
Eq.~(\ref{eq:transmuttation}), the absence of a term constant with
$\Delta N$ confirms that, as assumed in the random plane wave
approach, there are at this level of approximation no correlations
amongst the $\Si$.  Thus, the variance of the $\Si$ are given by
\begin{equation} \label{eq:variancesemi}
\mbox{Var}({\cal S}_i) = {\kf {\cal L} \over 2 \pi^3} \times 
\left\langle \left(\lambda(\theta) -  \langle
  \lambda\right\rangle_\theta  \right)^2 \rangle_\theta 
\end{equation}
with
\begin{equation} \label{eq:1/2classics}
\left\langle \left(\lambda(\theta) -  \langle
  \lambda\right\rangle_\theta  \right)^2 \rangle_\theta
= \left\{
  \begin{array}{ll} 
(2 \ln 2 -1) -  \left(\frac{\pi}{2} -1 \right)^2 & \mbox{Dirichlet } \\
(2 \ln 2 -1) -  \left(\frac{\pi}{2} -1 \right)^2 + 4\left(\ln
  {\pi \kf {\cal A} \over 2{\cal L}} -{\pi\over 2}\right) \qquad &
\mbox{Neumann}  
\end{array}
\right. \; .
\end{equation}

\subsection{The periodic orbit spectrum}
\label{tpos}

\begin{figure}[t]
\PSImagx{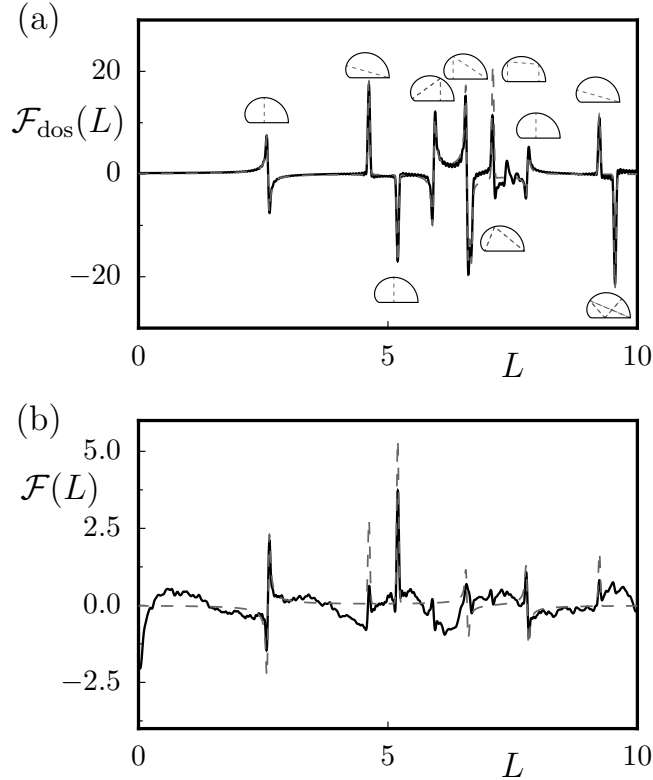}{8.6cm}
\caption{Fourier transform of the density of states and ${\cal S}_i - \left<{\cal S}_i \right> $ of the cardioid billiard using Dirichlet boundary conditions.  In (a), the solid line is for the density of states using the first 2000 odd parity levels of the cardioid billiard.  The dashed line is for the density of states using the corresponding form of the Gutzwiller trace formula. In (b), the solid line is the Fourier transform of ${\cal S}_i - \left<{\cal S}_i \right> $ for the first 2000 ${\cal S}_i$ and the dashed line is the result of Eq.~(\ref{eq:semiresult}) calculated for the shortest periodic orbits as the one shown in Fig.~\ref{fig:billiards} and the insets (along with their retracings).  The agreement is quite good.}
\label{fig:fouriertransform}
\end{figure}

Beyond the Var[$\Si$], which here characterizes the
universal (long time) behavior of the system under consideration, the
semiclassical treatment developed in the previous subsection provides
information on system specific quantities.  In particular, it
makes it possible to address phenomenon related to shorter time
dynamics, and thus quantum mechanically, to longer energy range.  For
example, Eq.~(\ref{eq:semiresult}) can be used directly to compute the
Fourier transform of the $\Si$.  Interestingly,
Eq.~(\ref{eq:semiresult}) has a structure very similar to that of
the density of states Eq.~(\ref{eq:rho_osc}).  This gives a
simple and striking prediction, namely that the Fourier transform of
the $\Si$ will display peaks at the same locations and with the same
shapes as the Fourier transform of $\rho(E)$ (up to the transformation
$i \leftrightarrow E$), and will be simply scaled by factors which
depend only on the lengths of the orbits and on their angles of incidences
$\{\theta_l\}$ at the various places where they bounce along the boundary.  
In Fig.~\ref{fig:fouriertransform}(a), the Fourier transform of the cardioid billiard density of states is shown in comparison with the Fourier transform of the $\{{\cal S}_i\}$ displayed in Fig.~\ref{fig:fouriertransform}(b).  The peaks are in precisely the same positions and their shapes are similar, but the amplitudes differ as expected; as a parenthetical remark, for technical reasons the Fourier transform of the $\{{\cal S}_i\}$ uses a slightly different Fourier transform than the density of states (effectively divided by the wave vector), which is denoted by a subscript in the remaining figures, but this has no effect on the overall  discussion of the physics involved.   In addition, the prediction of Eq.~(\ref{eq:semiresult}) for the shortest periodic orbits and their retracings is shown.  The predicted amplitudes for the $\{{\cal S}_i\}$ are reasonably close, although perhaps slightly too large by 30-50\%.  Otherwise, the agreement with Eq.~(\ref{eq:semiresult}) is excellent.  The excess in the prediction for short orbits is curious because if the predicted amplitudes for all of the orbits were too large, it should be found that the prediction of the variance is slightly too large instead of a bit too small  (say factor of two) as in this case.  We have checked that in fact, the predictions for long orbits, which dominate the calculation of the variance of the $\{{\cal S}_i\}$, are indeed a bit too small, the opposite of the short orbits.  Why it has turned out this way for this particular example remains for future consideration.

As a last remark, note that the tendency for long orbits to
explore uniformly the phase space implies both that $\sin \theta_l$
is distributed uniformly and that the mean length between bounces
$\ell_\gamma / n_\gamma$ for a given orbit $\gamma$ can be identified
with $\bar d$, the full system average distance between bounce.  This is
what made it possible to apply Eq.~(\ref{eq:savesourday}) and to
cancel the cross-terms between various fixed points of the same orbit
in Eq.~(\ref{eq:nocrosscor}).  Had the term proportional to
$n_\gamma^2$ in Eq.~(\ref{eq:nocrosscor}) not been zero, it would
have given rise to a contribution parametrically larger (in $\hbar$)
than the computed one [$\sim (\kf {\cal L})^2$ instead of 
$(\kf {\cal L})$].  Short orbits, for which the cancelation of cross-terms done in Eq.~(\ref{eq:nocrosscor}) cannot be applied may therefore have a stronger influence on the fluctuations of the $\Si$'s than what
might be naively expected from Eq.~(\ref{eq:varSN1}).

\subsection{The basic distinctions in the two theoretical approaches} 

Interestingly enough, the expressions in Eq.~(\ref{eq:1/2classics})
are exactly the same results as the random 
plane wave approach Eqs.~(\ref{eq:VarDir}-\ref{eq:VarNeu}), except
for two differences.  First,  the mean square $\langle
\lambda^2(\theta) \rangle_\theta$ has 
been replaced by the variance of $\lambda(\theta)$, giving now a much
better agreement with the billiard results (see Fig.\ref{fig:varianceSi}). 
Second there is a factor two difference in the prefactor.

\subsubsection{Proper normalization}

The replacement of the mean square by the variance can be
related to the lack of proper normalization of the wave function in
the random plane wave model.  Indeed, fluctuations of the local
density of states $\nu(\br)$ can either imply fluctuations of the
wavefunction probabilities $|\Psi(\br)|^2$, which have to integrate
to zero because of the wavefunction normalization, or fluctuations of
the total density of states for the part which survives the integration
over space.  The role of the term proportional to $\overline{\rho_{\rm
  osc}(E)}_{\Delta E}$ in the right hand side of Eq.~(\ref{eq:wf1})
can therefore be understood as ensuring the normalization of the
eigenfunctions.  As this term is precisely the one giving rise to the
contribution proportional to $ \left\langle
  \lambda\right\rangle_\theta^2 $, it turns out that in the
semiclassical calculation, proper normalization of the eigenfunction is
what generates the variance of $\lambda(\theta)$ rather than
its mean square.  Since the random plane wave model used here imposes
normalization on average rather than for each individual
eigenfunction, this contribution is necessarily missing there.  A modified version of the random plane wave model in which normalization is better enforced~\cite{Urbina07,Kaplan08} should,
however, properly address this issue.

\subsubsection{Dynamical correlations}

The factor two difference in the prefactor (or conversely, the fact
that except for this factor two and the normalization effect, the random plane wave
and the semiclassical expressions are identical), although less important from
a quantitative point of view, is however puzzling enough to
deserve further discussion.  To focus better on the main point, consider two
simplifications of the problem under discussion.  First, 
assume as understood the issue of eigenfunction normalization, and consider below only the contribution from the Green function (i.e., ignore density of states fluctuations).  Second,
consider that the procedure used to extract the variance $\varSi$ from the locally smoothed quantity $\overline{\cal S}_{\Delta N}$ [see Eq.~(\ref{eq:transmuttation})] is equivalent
to the effective rule according to which the various quantities under
consideration should be smoothed over an energy window of width $\Delta$
(so that $\Delta N = 1$).

Having this local smoothing in mind, and ignoring for the moment the
fluctuations of the density of states (i.e. assuming there is exactly
one state in each interval $\delta$) gives
\begin{eqnarray}
\overline{\left[ \sum_\kappa \Psi_\kappa(\br') \Psi^*_\kappa(\br')\right]}_\Delta  & = & \frac{1}{\Delta}   \Psi_\kappa(\br') \Psi^*_\kappa(\br') \\
& \simeq & -  \frac{1}{\pi} {\rm Im} \left [G^R(\br',\br',E) \right]
\; .
\end{eqnarray}
 Close to some reference point $\br$ and not considering yet the
 proximity of a boundary, this gives for the oscillating part of the
 wavefunction probability
 \begin{equation} {\cal A}  |\Psi(\br')|^2_{\rm osc}= -
   \frac{1}{\pi\nu_W}  \left[ \sum_{\mu:\br \to \br} A_\mu \exp
     \left({(\bp^f_\mu - \bp^i_\mu) \br'} \right) + c.c \right]\; .
\end{equation}
Above, the sum runs over all closed trajectories $\mu$ starting and ending
on the reference point $\br$, with initial and final momenta $\bp^i_\mu$
and $\bp^f_\mu$, time of travel $\tau_\mu$, and 
\begin{equation}
A_\mu = \frac{1}{i\hbar} \frac{1}{2i\pi\hbar}
\frac{1}{
    \sqrt{|\dot x_\mu \dot x'_\mu m_{12,\mu}|}} 
\exp\left[i \frac{S_\mu(\br,\br)}{ \hbar}-i\nu_\mu\frac{\pi}{2}
\right] 
\;
\end{equation}

In the semiclassical calculations of section~\ref{sec:gutzwiller}, it
is taken into account that as the integration over space is
performed, closed trajectories are continuously deformed, and in
particular the initial and final momenta $\bp^i_\mu$ and  $\bp^f_\mu$
vary.  As a consequence the dominant contributions, which 
correspond to nearly periodic trajectories (to within a bounce off the
billiard boundary in this particular calculation), can be 
understood as arising from the neighborhood of periodic orbits,
leading to the periodic orbit sum Eq.~(\ref{eq:gutz2}).  The
calculation of the variance is then done using the Hannay-Ozorio de
Almeida sum rule Eq.~(\ref{eq:HannayOzorio}).

Consider that if the dynamical correlations, i.e.~variations of the orbital properties (initial, final momenta $\bp^i_\mu$,  $\bp^f_\mu$, prefactor $A_\mu$, etc..) are neglected, the semiclassical expression for  $\Psi^*(\br') \Psi(\br')$ greatly resembles the random plane
wave model.  Indeed, if long orbits are dominant:

i) initial and final momenta $\bp^i_\mu$ and  $\bp^f_\mu$ are
independent and uniformly cover the energy surface; i.e.~the model
can be taken as a ``random pair of plane waves'' model.

ii) applying the diagonal approximation in the semiclassical calculation
amounts to $\overline{A^*_\mu A_{\mu^\prime}} \propto \delta_{\mu\mu^\prime} $ or, if the system is time reversal invariant $\mu$ and $\mu^\prime$ are related through time reversal invariance.  Near a boundary, the correlations are included between the trajectories related to one another by a bounce off the boundary.

iii) although the $A_{\mu}$ are not Gaussian distributed, the fact that
the number of trajectories is extremely large for long orbit makes it
possible to use a central limit theorem, implying that only the
variance of these quantities are relevant (and that one can as well
consider them as Gaussian).

iv) the variance of the $A_{\mu}$ is constrained by the sum rule valid
for closed orbits (a slightly different rule than the Hannay-Ozorio de
Almeida sum rule used for periodic orbits)~\cite{Argaman96}, 
\begin{equation}
\sum_\mu |A_\mu|^2 \delta(\tau-\tau_\mu) = \frac{2\pi}{\hbar} \nu_W P_{\rm
  cl}(\br,\br,\tau) 
\end{equation}
where for long orbits in billiards, the probalitity of return $P_{\rm
  cl}(\br,\br,\tau)$ can be taken uniform and equal to $1/{\cal A}$.
Properly carrying out the smoothing on the range $\Delta$ produce the damping factor ${\rm sinc}\left(\frac{\tau_\mu  \Delta}{2 \hbar} \right) $ of Eq.~(\ref{eq:damping}), so that 
\begin{equation} \label{eq:sumruleA}
\sum_\mu |A_\mu|^2  = \frac{2\pi}{\hbar} \frac{\nu_W}{{\cal A}} 
\int_0^\infty {\rm sinc}\left(\frac{\tau_\mu
      \Delta}{2 \hbar} \right) dt = 2 \pi^2 \nu_W^2 
\end{equation}
Thus, neglecting the spatial variations of the orbits properties which contain dynamical correlations, quite standard semiclassical approximations, namely the diagonal approximation and the assumption
that $P(\br,\br,\tau)$ is uniform, makes it possible to derive a ``pair
of random plane waves'' model, not completely identical to the
original random plane wave model, but similar in spirit.  It can be
shown furthermore than computing ${\rm Var}[S_i]$ under this
model gives exactly the same result as the random plane wave model.

Indeed, one can compute 
$ \left< \epsilon_i(\br_1) \epsilon_i(\br_2) \right> -\left< \epsilon_i(\br_1) \right> \left< \epsilon_i(\br_2) \right> = {\cal   A}^2 \left< |\psi_i(\br_1)|^2_{\rm osc}  |\psi_i(\br_2)|^2 _{\rm osc} \right>$ in this way [see Eq.~(\ref{eq:epsilon}))].  To be more precise, let the $\{A_\mu\}$ be uncorrelated unless the corresponding trajectories are time reversal symmetric, or related one to each other by a bounce off
the boundary of the billiard near the initial or final point of the
trajectory (i.e. $\left( (p^i_x)_{\mu'}, (p^i_y)_{\mu'} \right) = 
\left(\pm (p^i_x)_{\mu}, (p^i_y)_{\mu} \right )$ and 
  $\left( (p^f_x)_{\mu'}, (p^f_y)_{\mu'} \right ) = 
\left(\pm (p^f_x)_{\mu}, (p^f_y)_{\mu} \right )$.  If the reference
point $\br$ is taken on the boundary, and measuring the distance $x$
from the boundary, this amounts to taking $A_\mu$ equal for these
trajectories, giving 
\begin{eqnarray} 
{\cal  A}^2 &&
\left< \left(|\psi_i(\br_1)|^2\right)_{\rm osc} 
 \left(|\psi_i(\br_2)|^2\right)_{\rm osc} \right>
= \nonumber \\
&& \frac{2 \nu_W^2}{\pi^2} \sum_\mu |A_\mu|^2 \cos [(k_y^f -k_y^i)(y_1 -
y_2)]
{\rm cs}(p^f_x x_1/\hbar) {\rm cs}(p^i_x x_1/\hbar) {\rm cs}(p^f_x
x_2/\hbar) {\rm 
        cs}(p^i_x x_2/\hbar)  
\; ,\nonumber \\ \label{eq:elabet}
\end{eqnarray}
which, using the sum rule Eq.~(\ref{eq:sumruleA}) and inserting the
resulting wavefunction correlations in  Eq.~(\ref{eq:var}) gives
exactly Eq.~(\ref{eq:varianceexp}) derived with the random plane wave model.

To summarize, neglecting dynamical correlations in the semiclassical
approach generates a random pair of plane waves model that, in essence
is derived with usual approximations.  For the problem we consider
here, this model gives exactly the same result as the random plane
wave model.  The ``random pair of plane wave'' is however not ad hoc,
whereas the random plane wave model is.  This makes it possible to
discuss precisely what approximations have been made, and therefore in
what way we could expect the random model to differ from the purely
semiclassical treatment.  In particular we see that the interferences
between reflected wave at the boundary is treated in the same way in
both the semicalssical and the random approaches,  giving
rise to the same $\lambda(\theta)$ dependence.  On the other hand the
prefactor  is related, in the
semiclassical approach, to the way classical orbit with nearly
matching initial and final momenta are structured around periodic
orbit.  This aspect is completely ignored in the random models, and we
therefore cannot expect them to give exactly the correct
prefactor.

\section{Non-chaotic systems}
\label{sec:nfcs}

Included in the class of non-chaotic dynamical systems are three main
subclasses: i) the limiting case of integrable systems, all of whose
dynamics are regular; ii) near-integrable systems, characterized by
having classical perturbation theory generally work well in describing
its dynamics; and iii) mixed systems, which contain an intricate
mixture of both regular and chaotic dynamical regions in their phase
spaces.   Generally speaking, semiclassical theories and what is known
vary according to each subclass.  For example, trace formulae exist
for integrable~\cite{Berry76, Berry77a} and near-integrable
systems~\cite{Tomsovic95, Ullmo96}, but not for mixed systems.  In
fact, a proper treatment of semiclassical theory for mixed systems is
lacking.  However, for the purpose here of investigating the
properties of the set of $\{{\cal S}_i\}$, only the simplest level of
semiclassical theory is considered.  In other words, in regular
dynamical regions (whether from integrable, near-integrable, or mixed
systems), structures called tori are assumed to exist, which are
invariant manifolds under classical motion, and possible complications
from resonances, diffraction, or tunneling are ignored.  In chaotic
regions, only the complication of a family of marginally stable orbits
is considered beyond that which was already treated in the previous
section. 

Unlike chaotic dynamical regions in phase space, for regular regions
there are two possible overarching semiclassical approaches.  In the
first, particular tori quantize allowing the detailed evaluation of
${\cal S}_i$ for each eigenstate.  This is based on the
Einstein-Brillouin-Keller (EBK)
scheme~\cite{Einstein17,Brillouin26,Keller58}.  In the second, a
periodic orbit trace formula results from applying a Green function
approach, much like the chaotic case.  However, this Green function
approach is not given here since the information about each individual
${\cal S}_i$, the mean, and variance are already understandable
through the EBK approach. 

\subsection{Einstein-Brillouin-Keller quantization}
\label{sec:EBK}

\subsubsection{General expression for ${\cal S}_i$}

Continuing with two dimensional billiards, each torus is characterized by two
action variables $(J_1,J_2)$, and it is always possible to choose the
corresponding angles $(\varphi_1,\varphi_2)$ such that the
intersections of the torus with the boundary of
the billiard are parameterized as $\varphi_1 \! = f_\kappa(\varphi_2)$,
$\kappa = 1,\ldots,\kappa_{\rm max}$, with $\kappa_{\rm max}$ the
number of bounces on the boundary for the considered torus.

  An eigenstate $\Psi_i$ is constructed on a quantizing torus $\left[J_1 =
  2\pi \hbar (n_1^{(i)} + \sigma_1/4), J_2 = 2\pi \hbar (n_2^{(i)} + 
\sigma_2/4) \right]$ where $(\sigma_1,\sigma_2)$ are the Maslov indices  and can
be expressed as   
\begin{equation} 
\Psi_i({\bf r}) = \frac{1}{2\pi} \sum_\ell \sqrt{ \left| 
\frac{\partial (\varphi_1,\varphi_2)}{\partial (x,y)} \right|_\ell }
\exp\left( \frac{i}{\hbar}S_\ell(x,y) \right) \; .
\end{equation}
The sum runs over the various sheets of the torus projecting onto the
point ${\bf r} = (x,y)$ and $S_\ell(x,y)$ is the corresponding action
(including the Maslov phases).

Consider in greater detail, the neighborhood of the
$\varphi_1=f_\kappa(\varphi_2)$ boundary.  To further simplify the
discussion, assume that the torus has only two sheets (corresponding
to negative and positive $\varphi_1 - f_\kappa(\varphi_2)$) projecting
on any given point $(x,y)$ near this boundary.  The results derived
under this hypothesis apply in the general case, as is justified
below.  Adding the two $\varphi_1<f_\kappa(\varphi_2)$ and
$\varphi_1>f_\kappa(\varphi_2)$ contributions, and expanding the
action from the boundary as $S(x,y) = S(x\!=\!0,y) + p_x x$, generates
\begin{equation}  \label{eq:tildeSi_int}
\Psi(x,y) =\frac{1}{2\pi} \sqrt{ \left| 
\frac{\partial (\varphi_1,\varphi_2)}{\partial (x,y)} \right| }
\exp\left( \frac{i}{\hbar}S(x\!=\!0,y) \right) 2 {\rm cs} (p_x x
/ \hbar) \; ;
\end{equation}
assuming that the local variation of the Jacobian determinant in the
direction perpendicular to the boundary can be neglected.  Inserting
this expression into Eq.~(\ref{eq:Si}) we obtain
\begin{equation}
{\cal S}_i = i \left(1 \mp \frac{\cal L}{\kf {\cal A}} \right) \pm
\tilde {\cal S}_i
\end{equation}
with 
\begin{equation} \label{eq:calSi_one}
\tilde {\cal S}_i = \frac{i}{\pi^2} \int {\rm d}x {\rm d}y \left| 
\frac{\partial (\varphi_1,\varphi_2)}{\partial (x,y)} \right|
\frac{ J_1(2 \kf x) }{\kf x} {\rm cs}^2 (p_x x / \hbar) \; .
\end{equation}
If the torus has more than two sheets projecting onto the neighborhood
of the boundary (in which case Eq.~(\ref{eq:tildeSi_int}) involves a
sum), the rapidly oscillating phases, $\{\exp ( i S(x\!=\!0,y) /
\hbar)\}$, eliminate cross terms upon integration over $y$, and thus
the calculation of $\tilde {\cal S}_i$ would involve just a single
sum.

Changing the integration variables to
$(\varphi_2^0,\tau)$, with $\tau$ measuring
the time from the bounce on the boundary of the billiard and
$\varphi_2^0$ the angle $\varphi_2$ at that bounce, we can  further
simplify Eq.~(\ref{eq:calSi_one}).   Indeed 
\begin{eqnarray}
\varphi_2 & = & \varphi_2^0 + \omega_2 \tau \\
\varphi_1 & = & f_\kappa(\varphi_2^0)  + \omega_1 \tau \; ,
\end{eqnarray}
with $\omega_i = \partial H / \partial \varphi_i$ (i=1,2) the angular
frequencies, and   therefore the Jacobian can be expressed as 
$J = \left| \frac{\partial (\varphi_1,\varphi_2)}{\partial (\tau,\varphi_2^0)}
\right | = |\omega_1 - \omega_2 (df_\kappa/d\varphi^0_2)|$.  Noting furthermore
that $x(\varphi_2,\tau) = \vf \cos(\theta(\varphi_2^0)) \tau$, the
integral on the variable $\tau$ can be performed explicitly, giving
\begin{equation} \label{eq:intsi}
\tilde {\cal S}_i =  \frac{i}{\pi \kf \vf } 
\sum_{\kappa=1}^{\kappa_{\rm max}}
   \frac{ 1}{2 \pi} \int_0^{2\pi}  {\rm d}\varphi^0_2\
|\omega_1 -   \omega_2 (df_\kappa/d\varphi^0_2)|
\lambda\left[ \theta_\kappa(\varphi^0_2) \right] \; .
\end{equation}

\subsubsection{Circular billiard}

Beyond the orders of magnitude, the explicit evaluation of  the variance of
the set $\{{\cal S}_i\}$ for integrable systems is  very much
system dependent as it is usually not possible to make general
assumptions about the correlations between the various quantities involved
($\lambda(J_1,J_2,\theta_\kappa(\varphi^0_2)), \omega_{1,2}(J_1,J_2),
df_\kappa/d\varphi^0_2(J_1,J_2,\theta_\kappa(\varphi^0_2))$.  We
therefore consider now a specific system, namely the circular
billiard.

The computation of the  expression Eq.~(\ref{eq:intsi}) for this
billiard is made somewhat simpler because the angle $\theta$ at which
trajectories bounce off the boundary is a constant for a given torus,
and thus a function of the actions $(J_1,J_2)$ only.  Furthermore, 
for a given invariant torus, we can
construct the paths on which the action variables $(J_1,J_2)$ are
constructed as the cut of the torus in the radial direction (for
$J_1$), and the caustic (or any topologically equivalent path) for
$J_2$.  In this way, the angle $\varphi_2$ can be identified with the
angle of the polar coordinates, and the boundary of the billiard can
be taken as $\varphi_1=0$  (i.e. $f_{\kappa=\kappa_{\rm max} = 1}
\equiv 0$).  Furthermore the angular frequency $\omega_1(J_1,J_2)$ can be
identified with $2\pi/ t(J_1,J_2)$, with $t(J_1,J_2)$ the time between two
successive bounces for trajectories of the corresponding torus.  The
expression  Eq.~(\ref{eq:intsi}) thus takes the simple form
\begin{equation} \label{eq:circlesi}
\tilde {\cal S}_i =  \frac{2 i}{\kf \vf } 
\frac{\lambda\left[ \theta(J_1,J_2) \right]}{t(J_1,J_2)}
 \; .
\end{equation}

Quantizing tori sample uniformly the plane $(J_1,J_2)$, and therefore
statistical quantities such as average and variance should be computed
with the measure $dJ_1 dJ_2$. However, using that the change of
variable $(J_1,J_2,\varphi_1,\varphi_2) \to (E,\xi,\tau,p_\xi)$
introduced in appendix~\ref{app:meanl}  ($p_\xi \equiv \pf \sin \theta$)
is canonical, implying $dJ_1 dJ_2 d\varphi_1 d\varphi_2 = dE dp_\xi d\tau
d\xi$, and that for the circular billiard $\theta$ and $E$ depend only
on the action $(J_1,J_2)$, and $\tau$ and $\xi$ on the angles
$(\varphi_1,\varphi_2)$, one can write, in the neighborhood of the
Fermi energy $\Ef$
\begin{equation}
      dJ_1 dJ_2 \delta(E - \Ef) \propto t(J_1,J_2) d(\sin\theta) \; ,
\end{equation} 
and thus use the probability measure 
\begin{equation}
d P = \frac{t(\theta)}{\langle t(\theta) \rangle}  d\sin(\theta) \,
\end{equation} 
with $\langle T(\theta) \rangle$ given by Eq.~(\ref{eq:mean_t}) arising
from the normalization (given for allowed values of $\sin\theta$
  in the range $0 \le \sin\theta \le 1$).  

One can in this way recover in the particular case of the circular
billiard the general expression Eq.~(\ref{eq:simean}) of the mean
value $\langle {\cal S}_i \rangle$. Indeed
\begin{eqnarray}
\langle \tilde {\cal S}_i \rangle &=& \frac{2i}{\kf} \frac{\cal L}{
  \pi {\cal A}}  
\langle \lambda(\theta) \rangle_\theta \ \ \ {\rm and} \nonumber \\
\langle {\cal S}_i \rangle &=&  i \left(1 \mp \frac{\cal L}{ \kf
    {\cal A}} \right) \pm  2i\frac{\cal L}{\pi \kf   {\cal A}}
\left(\frac{\pi}{2} \pm 
1\right) = i \left(1+\frac{2 {\cal L}}{ \pi \kf {\cal A}}\right) \; .
\end{eqnarray}

Similarly,  the expression for the variance reduces to
\begin{eqnarray}
{\rm Var}\left[ {\cal S}_i\right] &=& \frac{4 i^2{\cal L}}{ \kf^2 \vf \pi {\cal A}}  \int_0^1 {\rm d}(\sin\theta) \frac{\lambda^2(\theta)}{T_1(\theta)} - \left[ \frac{i{\cal L}}{\kf  {\cal A}} \left(1 \pm \frac{2}{\pi}  \right)\right]^2 \nonumber \\
&=& \frac{4 i^2}{ \kf^2 \pi R^2}  \int_0^1 \frac{{\rm d}(\sin\theta)}{\cos\theta} \frac{(1\pm \sin\theta)^2}{\cos^2\theta} - \left[ \frac{2i}{\kf R} \left(1 \pm \frac{2}{\pi}  \right)\right]^2  \  .
\end{eqnarray}
As before, the constant for Dirichlet boundary conditions is quite
small, and the divergence of the Neumann boundary conditions increases
the order.  In this case, the effect is greater than logarithmic.
Using the same cutoff for the Neumann case as at the end of
Sect.~\ref{sec:fluctuations}, and that $i\approx {\cal A}\kf^2/4\pi$,
we obtain
\begin{equation}
{\rm Var}\left[ {\cal S}_i\right] = i \times \left\{ \begin{array}{ll}
\frac{2}{\pi} - \frac{1}{2} - \left(1- \frac{2}{\pi}  \right)^2
\approx 0.00457\qquad & {\rm Dirichlet} \\ 
-\frac{2}{\pi} - \frac{1}{2} + \pi^{1/2}i^{1/4} - \left(1- \frac{2}{\pi}  \right)^2 & {\rm Neumann}
\end{array}\right. \; .
\end{equation}
  Thus the variance
scales proportionally to $i$ for Dirichlet boundary conditions and
$i^{5/4}$ for Neumann boundary conditions.

\subsection{Bouncing ball modes in the stadium billiard}
\label{sect:bb}

Even for fully chaotic systems, it is possible to have a situation
where some (with vanishing measure) of the trajectories behave more
like those of integrable systems.  An example is 
provided by the bouncing ball orbits of the stadium billiard~\cite{Sieber93}.  Tanner~\cite{Tanner97} showed that for the
purposes of a semiclassical theory of eigenstates, the phase space in
the neighborhood of the bouncing ball orbits behaved much like an
island of regular motion, and that families of orbits that cannot be
taken as isolated contribute in essential ways.  This greatly
complicates the desire for a rigorous semiclassical theoretical
approach. Though these states can be thought of as EBK-like states
similar to those studied in Sect.~\ref{sec:EBK}, they do get connected
through diffractive terms to the chaotic states, and thus some of them
behave more like resonances rather than individual quantized states. 

A complete semiclassical description is beyond the scope of this
study.  Indeed, as mentioned in the beginning of Sect.~\ref{sec:nfcs},
as many dynamical system complications as possible are being neglected
here.  Instead of attempting a rigorous semiclassical theory for the
bouncing ball modes, a rough approximation is given instead.  To be
specific, we consider here the even-even symmetry states of a stadium
billiard with Dirichlet boundary conditions, or equivalently a
symmetry-reduced quarter stadium with Dirichlet boundary conditions on
the original boundaries and Neumann boundary conditions on the
symmetry lines. To start, consider the eigenstates of a rectangle with
one side length equivalent to the side length $L_s$ of the symmetry-reduced
quarter stadium and the other, the radius of curvature $R$.  These states
can be used to give an approximation to the bouncing ball modes.  In
essence, the bouncing ball modes with few nodes along the side length
(ignoring mixing into the chaotic states) vanish quickly upon entering
the quarter circular end-cap.  A quantization along the side length
direction with Dirichlet boundary conditions on the side entering the
end-cap and along the side length itself is a good starting point.
Since our calculations have been done for even-even symmetry
eigenstates, consider Neumann boundary conditions for the remaining
two sides.  The normalized states are given in cartesian coordinates
by
\begin{equation}
\left| \Psi_i (q_1,q_2) \right|^2 = \frac{4}{RL_s}
\cos^2\left(\frac{2m+1}{2L_s}\pi q_1  \right)
\cos^2\left(\frac{2n+1}{2R}\pi q_2 \right) 
\end{equation}
where the origin is the corner.  $m$ is a small integer, say $0,1,2,3$
or so, and most of the kinetic energy is in the other direction and so
$n$ is a large integer. 

To write the equation for the ${\cal S}_i$, the  coordinate system of
the state must be rotated and translated to the boundary coordinate
system used for the Friedel oscillations separately along the two
symmetry lines and the side length.  This gives three terms to
evaluate for the bouncing ball contributions, 
\begin{eqnarray} \label{eq:resultstad}
{\cal S}_i^{\rm  (bb)} &=& i\left(1 + \frac{{\cal L}_D-{\cal L}_N}{\kf {\cal
      A}} \right) + \frac{2i}{L_s} \int_0^\infty {\rm d}x
\frac{J_1(2\kf x)}{\kf x} \cos^2\left(\frac{2m+1}{2L_s}\pi x  \right)
\nonumber \\ 
&&\qquad \qquad + \frac{2i}{R} \int_0^\infty {\rm d}x \frac{J_1(2\kf
  x)}{\kf x}\left[ \cos^2\left(\frac{2n+1}{2R}\pi x \right)  +
  \sin^2\left(\frac{2n+1}{2R}\pi x \right)  \right] \nonumber \\ 
&=& i\left(1 + \frac{{\cal L}_D-{\cal L}_N}{\kf {\cal A}} \right) +
\frac{2i}{\kf R} + \frac{i}{\kf L_s}\left(1 + \sqrt{1-\left[
      \frac{(2m+1)\pi}{4\kf L_s} \right]^2}  \right) \nonumber \\ 
{\cal S}_i^{\rm (bb)} - \left\langle {\cal S}_i \right\rangle &=& \frac{i}{\kf {\cal A}}\left[{\cal L}_D-{\cal L}_N - \frac{2{\cal L}}{\pi} + 2{\cal A}\left(\frac{1}{R} + \frac{1}{L_s}\right)\right]
\end{eqnarray}
where the straightforward integrals over the $y$ coordinates have been
evaluated before writing the first expression, and ${\cal L}_D = L_s +
\pi R/2$ and ${\cal L}_N = L_s + 2 R$ are the perimeter
lengths with Dirichlet and Neumann boundary conditions respectively.
In the last expression, the overall mean Eq.~(\ref{eq:simean}) is
subtracted and the square root reduces to unity since bouncing balls
with significant momentum toward and away from the end-cap do not
exist.

There are a number of interesting consequences of
Eq.~(\ref{eq:resultstad}).  The last line captures the scale of the
deviation from the mean.  Putting in the parameters used for the
stadium calculations of Figs.~(\ref{fig:mean},\ref{fig:varianceSi})
($R=L_s=1$) generates a growing expected deviation from the mean that
hits about $30$ for $i=2000$.  This result also implies that ${\cal
  S}_i$ for the non-bouncing ball modes fluctuate about a negative
bias given by  
\begin{equation}
\label{eq:negbias}
{\cal S}_i^{\rm (non \, bb)} - \left\langle {\cal S}_i \right\rangle =
\frac{-f_{\rm bb}}{1-f_{\rm bb}} \frac{i}{\kf {\cal A}}\left[{\cal
    L}_D-{\cal L}_N - \frac{2{\cal L}}{\pi} + 2{\cal
    A}\left(\frac{1}{R} + \frac{1}{L_s}\right)\right] \ . 
\end{equation}
where the fraction of bouncing ball modes is denoted $f_{\rm bb}$.
The best known estimate of $f_{\rm bb}$ for the stadium billiard to
our knowledge comes from Tanner~\cite{Tanner97}, which implies that
$f_{\rm bb} \approx \gamma \left[\frac{2}{\pi(\gamma +
    \pi/4)}\right]^{3/4} i^{-1/4} $; we have introduced the ratio
$\gamma = L_s/R$ and used that $4\pi i ={\cal A}\kf^2$.  For other
systems with bouncing ball modes, their fraction may scale differently~\cite{Baecker97}. The results of
Eqs.~(\ref{eq:resultstad},\ref{eq:negbias}) are given by the dashed and solid
lines in Fig.~\ref{fig:mean}(b).  The dashed (upper) line
crosses right through the neighborhood of the peak values.  For such a
rough approximation, Eq.~(\ref{eq:resultstad}) is quite good.  In
addition, the solid (lower) line captures the negative bias implied for
the non-bouncing ball modes quite well.

Second, to the level of approximation here, any of the bouncing ball
modes contribute fairly equally locally in energy (determined by the
quantum number $n$); i.e.~there is almost no dependence on sequence
number $m$.  Although, the peak values from the bouncing balls do not
appear to be constant in the figure, presumably due to weak admixtures
of chaotic states, this feature greatly simplifies a calculation of
the fluctuations.  The variance can be inferred by noting the
following: i) the square of the bouncing ball deviation from the mean
multiplied by their fraction denoted $f_{\rm bb}$ gives their
contribution; and ii) the contribution of the remaining $1-f_{\rm bb}$
non-bouncing ball modes can be taken as the square of the average
amount they must each be in deficit of the mean plus the [sub-leading]
variance from the chaotic system results.  The combined consequences
lead to the expression
\begin{eqnarray}\label{eq:variancefb}
{\rm Var}\left[{\cal S}_i\right] &=& \frac{f_{\rm bb}}{1-f_{\rm bb}}
\frac{i}{4\pi {\cal A}}\left[{\cal L}_D-{\cal L}_N - \frac{2{\cal
      L}}{\pi} + 2{\cal A}\left(\frac{1}{R} +
    \frac{1}{L_s}\right)\right]^2 +  {\rm Var}\left[{\cal
    S}_i\right]_{\rm chaotic} \quad {\rm with}  \nonumber \\ 
{\rm Var}\left[{\cal S}_i\right]_{\rm chaotic} &=&  {\kf {\cal L} \over 2
  \pi^3}  \left[ (2 \ln 2 -1) -  \left(\frac{\pi}{2} -1
  \right)^2\right]  + {2\kf {\cal L}_N \over \pi^3} \left(\ln {\pi \kf
    {\cal A} \over 2{\cal L}} -{\pi\over 2}\right) \nonumber \\ 
\end{eqnarray}
Taking account of the decreasing fraction of bouncing ball modes with
increasing $i$, the variance scales as $i^{3/4}$.  See
Fig.~\ref{fig:varianceSi} for a comparison of this formula with ${\rm
  Var}[{\cal S}_i ] $ for the even-even eigenstates of the stadium
billiard.  Again, the rudimentary approach here captures the main
behavior fairly well.

To conclude this subsection on the bouncing ball orbits, a few remarks
are in order.  First, we stress that although the classical dynamics
of the stadium billiard is, mathematically speaking, purely chaotic,
as far as ${\cal S}_i$ statistics are concerned the existence of the
marginally stable bouncing ball family makes this system behave very
much like an integrable billiard. In particular the scale of the
fluctuations are order of magnitude larger than for the ``genuine''
chaotic billiard considered in sections~\ref{sec:rpw} and
\ref{sec:gutzwiller}.  Second, because of the presence of two classes of
states with drastically different properties, the covariance amongst
the ${\cal S}_i$ is, again in contrast with genuine chaotic systems,
non zero.  Furthermore, both the variance and the covariance
show an energy dependent structure which complicates
significantly the extraction of these quantities from the locally
smoothed $\overline{{\cal S}_i}_{\Delta N}$ as was done in
section~\ref{sec:gutzwiller}.

\begin{figure}[t]
\PSImagx{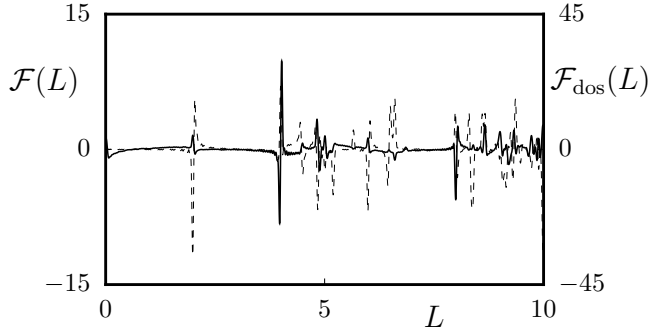}{8.6cm}
\caption{Fourier transform of ${\cal S}_i - \left<{\cal S}_i \right> $
  and the density of states from the even-even stadium eigenstates
  using Dirichlet boundary conditions.  The dashed line is for the
  density of states (which has a different vertical scale shown 
  at the right side).} 
\label{fig:ftstad}
\end{figure}

As a final remark, although the periodic orbit formula is not derived
here, Fig.~\ref{fig:ftstad} is shown for completeness.  As in 
Fig.~\ref{fig:fouriertransform}, the results are
compared between the density of states and the $\{ {\cal S}_i\}$.  Again, the
peaks are in the same positions, i.e.~those determined by periodic
orbits, but with differing amplitudes.  The importance of the bouncing
ball modes is quite visible. 

\section{Discussion}
\label{sec:discussion}

The mean and variance of the quantities $\{{\cal S}_i\}$ treated in~\cite{Tomsovic08} and in greater detail in this paper have been introduced as examples of a new, and non-local, class of statistical measures with physical relevance; the $\{{\cal S}_i\}$ are connected to the addition spectrum of quantum dots.  By no means should they be taken as the only possible measures representing this class.  Indeed, in recent works on finite-size fluctuation properties in ultracold Fermi gasses~\cite{Olofsson08, Garcia08}, involving fluctuations in the Bardeen-Cooper-Schrieffer pairing gap, a quantity was introduced for the order parameter that is similar to the $\{{\cal S}_i\}$, though with additional complications.  
Not surprisingly, several common features are found for both quantities.  As the fluctuations are dominated by a term arising from the interplay of the Friedel and eigenstate oscillations, there is a significant decrease in fluctuation magnitude due to Dirichlet boundary conditions, either through reduction or even vanishing of the prefactor of the leading term in $\kf L$.  In the case of the $\{{\cal S}_i\}$, the prefactor is decreased by more than an order of magnitude with respect to the prefactor for Neumann boundary conditions.  Another critical feature is the role of dynamics in the scale of the fluctuations.  Chaotic dynamical systems lead to a fluctuation scale of lower order in $\kf L$ than integrable or mixed dynamical systems.  Again, this leads to a fluctuation scale decreased by an order of magnitude or more.  The decrease in scale for chaotic systems can be traced to an ergodic nature of the individual eigenstates.  Conversely, the much larger fluctuation scale for integrable and mixed phase space systems, suggests the possibility of new physics associated with more regular dynamics or the possibility of measurements that can be used to deduce information about the dynamics.

The $i^{1/2} \propto \kf {\cal A}/{\cal L}$ (or $ i^{1/2}\ln i$) dependence for chaotic systems, and faster-growing dependence for integrable systems, implies that the fluctuations embodied in ${\rm Var}[{\cal S}_i]$ in fact grow with the size of the system, becoming eventually larger
than of order unity.  This implies that the corresponding residual interaction
contributions will in that case become larger than the mean level
spacing $\Delta$.  In other words, since $\Delta$ is the energy scale
set by the one particle energies, the fluctuations in the residual
interactions may become large enough that they generate a modification
in the ground state orbital occupation number, and more generally
reach a point where a first order perturbation treatment of the
interactions is not adequate.

Returning to the two theoretical approaches included here -- a random plane wave model for chaotic systems and semiclassical theory, whatever the dynamics -- we have seen that the basic random plane wave model is intrinsically less powerful than semiclassical theory, but on the other hand, it is technically much simpler to implement.  Surprisingly, given the excellent results it generates in other contexts, the random plane wave model here displays a couple of significant faults.  The most simple to track down is the effect of having its normalization be across the ensemble as opposed to the individual eigenstates.  For chaotic systems, this led directly to the replacement of the mean square of $\lambda(\theta)$ by its variance.  Inclusion of the variance improved the theory considerably as the mean square largely overestimated the prefactor.  In addition, the proper treatment of dynamical correlations in the semiclassical theory led to a factor two increase in the prefactor constant.  Truth be told, for the cardioid billiard example, the results seem to agree better without the factor two (see Fig.~\ref{fig:fouriertransform}), but a concerted search for an error in the semiclassical calculation never resulted in its removal.

Finally, it is important to be aware of some consequences of decompositions such as given in Eq.~(\ref{eq:nre2}).  The separation of average and fluctuating parts of a non-local statistical measure may not be the full story from a physical perspective.  In fact, this is the case for the $\{{\cal S}_i\}$ treated here.  As noted earlier, the mean of ${\cal S}_i$ does not lead to any modifcation of  the ground state occupations numbers as its effecs gets cancelled.
   On the other hand, the fluctuating part of ${\cal S}_i$ does affect the ground state, but there are two components.  The leading order fluctuations that come from the use of the term $N_{\rm sec}(\br;E_i^+)$ in the fluctuation expressions is essentially a mean field effect, analogous to scrambling.  These fluctuations, if large enough, can reorder the filling of the single particle levels.  They do not however lead to high spin states or other unusual behaviors.  The remaining term $\delta N(\br;E_i^+)$ is responsible for exotic physics in those cases where it is sufficiently large.  As the focus throughout this paper was on developing the theory of the non-local statistical measures themselves, the leading behaviors have been emphasized, which though dominant, are not necessarily the only ones that deserve to be considered - that depends on the physical context (other types of problems exist, such as the fluctuations of superconducting gap mentioned ealier, where the dominant terms in the relevant non-local statistical measures do contain all the important physics).  Therefore, revisiting the fluctuation-fluctuation term involving $\delta N(\br;E_i^+)$ will sometimes be important, but is left for future work.

\appendix

\section{Second order terms in the average properties of the ${\cal S}_i$}
\label{app:secondorder}

There are second order terms in the computation of the average properties of the $S_i$ coming from the boundary, which are given here.  Although, the other second order terms from the curvature and boundary discontinuities are not being derived and hence this calculation is incomplete, the numerical calculations necessary to isolate the average behavior before calculating the variance or covariance are improved by including them.  Therefore, an account is given here.

The decomposition of Eq.~(\ref{eq:nre2})  giving $\delta N(\br;E_i^+)$ implies
\begin{equation}
\delta N(\br;E_i^+) = \sum_{j\le i}  \left|\psi_j(\br)\right|^2  -
\left<\left|\psi_j(\br)\right|^2\right> \;.
\end{equation}
Substituting the relations from Eqs.~(\ref{eq:nre2},\ref{eq:psidef}) and integrating
the constant terms gives  
\begin{equation} \label{eq:Si2}
{\cal S}_i = i \pm \frac{1 }{ {\cal A}} \int {\rm d}\br \ \left[ \frac{i
    }{ \left( 1 \pm \frac{{\cal L}   }{ k {\cal A}} \right) }
  \frac{J_1(2 \kf x)}{\kf x}  \epsilon_i(\br) \pm {\cal A} \delta
  N(\br;E)\epsilon_i(\br)  \right] \;.
\end{equation}
The last term merits some discussion:  its leading behavior is seen to
be two orders weaker than the overall expression and is
straightforward to evaluate because only its leading contribution is
required.  Under the operation of taking the expectation value, the
only surviving term is 
\begin{equation}
\label{eq:expect}
\left< \delta N(\br;E)\epsilon_i(\br) \right> = {\cal A}\left<
  \left[\left|\psi_i(\br)\right|^2  -
    \left<\left|\psi_i(\br)\right|^2\right>\right]^2 \right>  \; , 
\end{equation}
so that after integration over space, this is essentially equivalent to $\langle
M_{ii} \rangle - \langle M_{ij} \rangle $. Indeed, the local Gaussian
random behavior is uncorrelated from state-to-state in the random
plane wave model so that the only surviving term comes from the
$i^{th}$ state with itself, all others vanishing. Locally, before
squaring and taking the expectation value, the expression on the r.h.s.~of Eq.~(\ref{eq:expect}) can be thought of as being like the square of a zero-mean,
unit-variance Gaussian random variable with unity subtracted.
However, it does have a variance, which is position-dependent and given by
the right-hand-side of Eq.~(\ref{eq:envelope}); i.e.\ the r.h.s.~acts
as an envelope.  Inside of the billiard (excluding the semiclassically vanishing
boundary region), its value is however constant and equal to
the inverse of $\cal A$ to leading order.  This generates a constant,
equal to two, after integration.  Therefore, the expectation value of
${\cal S}_i$ is approximately
\begin{eqnarray}
\label{eq:simean2}
\left<{\cal S}_i \right> &=& i + 2 \pm \frac{i}{ {\cal
    A}\left( 1 \pm \frac{{\cal L}   }{ \kf  {\cal A}} \right)} \int {\rm
  d}\br \  \frac{J_1(2\kf x) }{ \kf x}  \left<\epsilon_i(\br)\right>
\nonumber \\ 
&=& i + 2 + \frac{ i}{ {\cal A}\left( 1 \pm \frac{{\cal L}
      }{ \kf  {\cal A}} \right)\left( 1 \pm \frac{{\cal L}   }{ 2\kf  {\cal
        A}} \right)} \int {\rm d}\br \  \frac{J_1(2\kf x) }{ \kf x}  \left[
  J_0(2\kf x)  - \frac{{\cal L} }{ 2 \kf  {\cal A}} \right] \nonumber \\ 
&=& i \left( 1 +  \frac{{\cal L}   }{ \kf  {\cal A}} \frac{1}{  \left( 1 \pm
     \frac {{\cal L}   }{ \kf  {\cal A}} \right)\left( 1 \pm \frac{{\cal L}
        }{ 2\kf  {\cal A}} \right)} \left[ \frac{2 }{\pi } - \frac{{\cal L}
      }{2\kf  {\cal A}}\right] \right) + 2 \nonumber
\\ 
&=& i \left[ 1+  \frac{2{\cal L} }{\pi \kf  {\cal A}} \mp  \left( \frac{3
      }{\pi } \pm  \frac{1 }{ 2}\right) \frac{{\cal L}^2 }{ \kf ^2 {\cal
      A}^2}\right] + 2 
\end{eqnarray}
where we choose $\kf =\sqrt{2mE_i}/\hbar$.  Interestingly enough, if a length ${\cal L}_D$ of the boundary follows Dirichlet  conditions and a length ${\cal L}_N$ follows Neumann conditions, it is not correct just to make the substitution $\pm{\cal L}\rightarrow {\cal L}_N-{\cal L}_D$.  Rather, the above expression becomes
\begin{equation}
\left<{\cal S}_i \right>  =  i \left[ 1+  \frac{2{\cal L} }{ \pi \kf  {\cal A}} -  \left( \frac{3{\cal L}  }{\pi } +
  \frac{{\cal L}_N - {\cal L}_D  }{ 2}\right) \frac{{\cal L}_N - {\cal L}_D  }{ \kf ^2 {\cal A}^2}\right] + 2
\end{equation}
after redoing the algebra.  The distinction arises because some of the correction terms depend on the sign of the boundary conditions, whereas other correction terms depend on the sign squared.

\section{Mean length of a trajectory between two successive bounces}
\label{app:meanl}

This appendix briefly rederives Eq.~(\ref{eq:meanl}), which gives
the mean length of a trajectory between two successive bounces off the
boundary of a billiard. While this is a well-known result 
(see~\cite{Chernov97} and references therein), several equations
used in the derivation are needed in the main text.
The result can actually be obtained by computing in two different ways the
energy surface volume of the billiard
\begin{equation}
V(E) = \int d\bp d\br \delta(E - H(\br,\bp)) \; ,
\end{equation}
with $H(\bp,\br) = \bp^2/2m + V(\br)$, and $V(\br) = 0$ inside the
billiard and $\infty$ outside.  The first way is to perform this integral with the original coordinates
$(\br,\bp)$, giving  $V(E) = 2 \pi m {\cal A}$.

The second way to perform this integral is to use another set of
coordinates, constructed as follows.  Any point $(\br,\bp)$ 
of the billiard's phase space, can be considered as belonging to a
trajectory which has last bounced off the boundary a time $\tau$ ago at a
location on the boundary labeled by the curvilinear abscissa $\xi$.
Denote $\br_0(\xi)$ the corresponding point on the boundary and
introduce the action
\begin{equation}
S(r_1,r_2,\tau,\xi) \df \int_{\br_0(\xi)}^{\br=(r_1,r_2)} L(\tau') d\tau'
\end{equation}
with  $L = \bp \dot \br - H$ the Lagrangian function.  Since $\partial
S/\partial \br = \bp$, $S(r_1,r_2,\tau,\xi)$ can be used as the
generating function of the canonical transformation  
\begin{equation}  \label{eq:cantrans}
(\br,\bp) \longrightarrow ({\bf Q},{\bf P})
\end{equation}
with ${\bf Q} = (\xi,\tau)$.  The new momentum coordinates are thus given
by
\begin{eqnarray*}
P_1 & = & - \frac{\partial S}{\partial Q_1} = -\frac{\partial
  S}{\partial \tau} = -E \\
P_2 & = & - \frac{\partial S}{\partial Q_2} = -\frac{\partial
  S}{\partial \xi} = p_\xi 
\end{eqnarray*}
with  $p_\xi$ the projection of the momentum $\bp$ on the direction
parallel to the boundary at $\xi$.

As Eq.~(\ref{eq:cantrans}) is a canonical transformation, $d\br
d\bp = d{\bf Q}d{\bf P}$ and the energy surface
volume is 
\begin{equation}
V(E) = \int d\xi dp_\xi dE d\tau  \delta(E - H) \; .
\end{equation}
Performing the straightforward integration
over energy, and noting that for given  $(\xi,p_\xi)$ the integral
over $d\tau$ yields the time of travel $t(\xi,p_\xi)$ of the
corresponding trajectory between $\xi$ and the following bounce gives
\begin{equation}
V(E) = \int d\xi  dp_\xi t(\xi,p_\xi)  = 2 p {\cal L} \bar t \; ,
\end{equation}
with $p = \sqrt{2mE}$ and $\bar t$ the mean time of travel between two
successive bounces.  Identifying this expression with the one obtained
using the original coordinate system gives
\begin{equation} \label{eq:mean_t}
\frac{p}{m} \bar t = \frac{\pi {\cal A}}{{\cal L}} \, 
\end{equation}
and finally, using ${p}/{m} = |\dot \br|$, one finds Eq.~(\ref{eq:meanl}).

\section{Normalization corrections to Eq.~(\ref{eq:epsi}) }
\label{app:norm}

Note that in Eq.~(\ref{eq:epsi}) only the terms which are of leading
order {\em near the boundary} have been kept.  Others have been
neglected which, though smaller near the boundary, are of the same
size as some of the terms kept when integrated over the full area of
the billiard.  In particular, as it is, Eq.~(\ref{eq:epsi}) gives the
normalization of the wavefunction only up to ${\cal L}/\kf {\cal A}$
corrections.  If however  $\rho_W(E) (1 \pm {\cal L}/\kf {\cal A})$ is
used rather than $\rho_W(E)$ for the smooth part of the density of
states, Eq.~(\ref{eq:wf1}) is replaced by   
\begin{eqnarray}
 {\cal A} \overline{|\Psi_i(\br)|^2}_{\Delta N}  - 1 &=&  \pm J_0(2\kf x) \mp \frac{\cal L}{2 \kf {\cal A}} + \frac{{\cal L}^2}{4\kf^2 {\cal A}^2} - \frac{\cal L}{2\kf {\cal A}} J_0(2\kf x)  \nonumber \\
&& \frac{1}{{\cal A}^2 \nu_W^2}\ \overline{\rho^2_{\rm  osc}(E)}_{\Delta E} + \frac{1}{\pi {\cal A} \nu_W^2}\ \overline{\rho_{\rm  osc}(E)}_{\Delta E} {\rm Im}\  \overline{\tilde G_{\rm osc} (\br,\br,E)}_{\Delta E} \nonumber \\
&&- \frac{1 \mp  \frac{\cal L}{2 \kf {\cal A}}}{\pi \nu_W }  {\rm Im}\  \overline{\tilde G_{\rm osc} (\br,\br,E)}_{\Delta E} - \frac{1 \pm J_0(2\kf x) \mp \frac{\cal L}{\kf {\cal A}} }{{\cal A} \nu_W} \overline{\rho_{\rm  osc}(E)}_{\Delta E} \; .  \nonumber \\
\end{eqnarray}
Integration over the r.h.s gives precisely zero with each pair of terms from the beginning in order respectively canceling each other.  Thus, the eigenstate normalization is unity from the leading constant term on the l.h.s.  

\acknowledgments

We gratefully acknowledge discussions with Barbara Dietz-Pilatus and Thomas Friedrich.    One of us (S.T.) gratefully acknowledges support from  the U.S.~National Science Foundation
Grant  No.~PHY-055530, and one of us (A.B.) gratefully acknowledges support from the DFG 
under contract FOR760.


\end{document}